\documentclass{aa}  
\usepackage{natbib}
\usepackage{txfonts}
\usepackage{bm}
\bibpunct{(}{)}{;}{a}{}{,} 

\usepackage{graphicx}

\usepackage{hyperref}
\newcommand{\beq}{\begin{equation}}
\newcommand{\eeq}{\end{equation}}
\newcommand{\bea}{\begin{eqnarray}}
\newcommand{\eea}{\end{eqnarray}}

\newcommand{\cm}{\mathcal}
\newcommand{\mb}{\mathbf}

\newcommand{\kB}{k_\mathrm{B}}

\usepackage[normalem]{ulem}

\begin{document}

\title{A numerical code for the analysis of magnetic white-dwarf spectra that includes field effects on the chemical equilibrium}

\titlerunning{A numerical code for the analysis of magnetic white dwarfs}

\author{Mat\'ias Vera-Rueda and Ren\'e D. Rohrmann} 
 
\institute{Instituto de Ciencias Astron\'omicas, de la Tierra y del
Espacio (CONICET-UNSJ), Av. Espa\~na 1512 (sur), 5400 San Juan, Argentina}

\abstract{
We present a new magnetic-atmosphere model code for obtaining synthetic spectral fluxes of hydrogen-rich magnetic white dwarfs.
To date, observed spectra have been analyzed with models that neglect the magnetic field's effects on the atomic populations. In this work, we incorporate state-of-art theory in the evaluation of numerical densities of atoms, free electrons, and ions  in local thermodynamical equilibrium under the action of a magnetic field. The energy distribution of atoms is rigorously evaluated for arbitrary field strength. This energy pattern includes going from tightly bound states to metastable or truly bound, highly excited states embedded in the continuum, that is, over the first Landau level. 
Finite nuclear mass effects and the coupling between the internal atomic structure and the motion of the atom across the magnetic field are also considered.
Synthetic fluxes are generated with integrations of numerical solutions of polarized radiative transfer over the visible stellar disk using a spherical $t$-design method.  
The atmosphere code is tested with observations from the Sloan Digital Sky Survey for a group of known magnetic white dwarfs. Physical stellar parameters are obtained from least-squares fits to the observed energy distribution and compared with results of previous works.
We show that the use of zerofield ionization equilibrium in spectral analyses can lead to underestimated effective temperatures for highly magnetic white dwarfs.
} 


\maketitle
%

\section{Introduction}\label{s:intro}

Magnetic white dwarfs (MWDs) represent a very challenging field of research in astrophysics. \citet{blackett1947} suggested their existence several decades ago, but no evidence was found for a number of years. 
The first known MWD, Grw$+70^\circ 8247$, was identified by its strong circular polarization in the continuum spectrum \citep{kemp1970a}. 
By the late 1970s, sixteen other MWDs, including twelve isolated objects and four in binary systems, were recognized \citep{angel1978}.
Around the mid 1990s, about 40 MWDs with field strengths greater than 1 MG ($10^6$~gauss) were studied   \citep{schmidt1995}.
With the arrival of larger surveys, mainly the Sloan Digital Sky Survey (SDSS), the number of known MWDs has grown to over 600 isolated stars and about 200 objects in interacting binary systems \citep{ferrario2020}. About 800 MWDs have recently been analyzed in a comprehensive study \citep{amorim2023}.

Evidence of magnetic fields on the surface of white dwarfs has been deduced from the detection of broad-band circular polarization  \citep[e.g.][]{kemp1970a,angel1971, west1989, berdyugin2022}, cyclotron features observed in optical and UV spectra \citep[e.g.][]{visvanathan1979, green1981, bailey1991}, 
 Zeeman splittings of spectral lines \citep[e.g.][]{angel1974, liebert1975, wickramasinghe1976, kepler2013}, and spectropolarimetric measurements \citep[e.g.][]{aznar2004, afanasev2018, landstreet2019}. 
In some cases, the observation of a stable oscillation period may also suggest a rotating magnetic star with localized magnetic structures on its surface \citep{katz1975, landi1976, dupuis2000, kilic2015}. Signatures of spot-like components have been found in a number of white dwarfs \citep[e.g.][]{shtol1997, maxted2000, valyavin2008, vornanen2012, brinkworth2013} including the first candidate MWD in any globular cluster \citep{pichardo2023}.

Studies reveal a field distribution with surface strengths in the $10^3$--$10^9$~G range \citep{kawka2020}. Most MWDs have been discovered via Zeeman identifications in the SDSS \citep{gansicke2002, schmidt2003, vanlandingham2005, kepler2013, amorim2023} and present field strengths exceeding 1~MG due to the limit of detectability at the spectral resolutions used and because the Zeeman splitting is greater than the Stark broadening for this field range \citep{chanmugam1992}.\footnote{Although the field strength changes across the stellar surface and smears the spectral features (e.g., line wavelength $\lambda$), its mean value can be deduced from the so-called stationary lines ($d\lambda/dB\approx 0$).} 
The Zeeman splitting becomes undetectable below $\approx50$~kG \citep{bagnulo2018}. 
Detection of weak fields, below 1~MG and near a few kilogauss, is possible with the observation of polarization in spectral line wings through spectropolarimetry surveys \citep{schmidt1995, aznar2004, kawka2007, valyavin2006, landstreet2012}.

Surface magnetic fields are present in roughly 10\% of the total white dwarf population \citep{liebert2003, hollands2015}. This proportion is also found in MWDs with weak fields \citep{jordan2007, landstreet2012}.
A higher incidence of magnetism was found in studies of nearly complete 13 pc (around $20$\%) and 20 pc ($12$\%) volume-limited samples \citep{kawka2007, holberg2016}. Nevertheless, magnetic properties of about 80\% of white dwarfs are still unknown \citep{valyavin2015}, and MWDs with fields under 0.1~MG remain to be discovered \citep{kawka2007}.

The origin of magnetic fields in white dwarfs is not well understood. They may arise from ({\it i}) fossil fields retained during the evolution of magnetic progenitors \citep{fontaine1973, angel1981, wickramasinghe2005}, ({\it ii}) a preceding evolutionary stage through a dynamo process or a convective mechanism \citep{ruderman1973, levy1974, kissin2015, cantiello2016}, ({\it iii}) a binary system evolution through different mechanisms \citep{wickramasinghe2000, tout2008, garcia2012, briggs2015}, or ({\it iv}) the cooling of an already formed white dwarf \citep{isern2017}. 
Mergers \citep{ garcia2012, briggs2015} or accretion \citep{tout2008, nordhaus2011} could explain why magnetic white dwarfs tend to be significantly more massive \citep[mean mass $\approx 0.8$~$M_\sun$,][]{liebert1988, kawka2007, kepler2013} than nonmagnetic degenerates \citep[$\approx 0.6$~$M_\sun$,][]{kleinman2013}. A considerable amount of work is required to understand the origin and properties of the fields and their incidence on the structure and evolution of these stars.

The interpretation of MWDs requires detailed atmosphere modeling. The first attempts to explain the radiation of MWDs were based on a graybody magnetoemission model \citep{kemp1970b, shipman1971, chanmugam1972}.
However, an appropriated analysis of the radiation spectrum emitted at the surface of such stars demands numerical approaches for solving the transfer of polarized radiation through a magnetized medium \citep{unno1956, beckers1969, hardorp1976, martin1979, nagendra1985} and improved evaluations of the emission and absorption processes of gases in the presence of magnetic fields \citep[e.g.,][]{garstang1977, henry1984, roesner1984, ruder1994, merani1995, zhao2007}.
With advances in the required input physics, considerable effort has been made to progressively obtain more realistic model atmospheres \citep{martin1979, wickramasinghe1979, odonoghue1980, nagendra1985, schmidt1986, wickramasinghe1988, jordan1989, jordan1992, euchner2002}. 

Although reasonable fits to the MWDs' spectra of light emission have been achieved, including comprehensive analyses of a great number of objects \citep{kulebi2009, amorim2023, hardy2023}, a number of simplifications and rough approximations remain to be addressed. In particular, the direct effects of the magnetic field on the hydrostatic structure of the atmosphere are neglected \citep{wickramasinghe1988, jordan1992}. Consequently, zerofield models are used to provide the pressure and temperature distributions of magnetic-atmosphere models, ignoring induced Lorentz forces \citep{landstreet1987} and likely deformations of the atmospheric geometry \citep{stepien1978, fendt2000}.
Furthermore, detailed and rigorous evaluations of a number of radiative processes are still required. In particular, continuum opacities arising from bound-free and free-free transitions are poorly approximated, and an appropriate theory of Stark broadening of spectral lines for arbitrary magnetic field is not available \citep[although some efforts are in progress; e.g.,][]{kieu2017,rosato2023}.

Similarly, detailed calculations of the ionization equilibrium and occupation numbers of atomic levels in magnetic fields have not been included in any model of magnetic white dwarfs. In fact, until recently, no reliable evaluation of chemical equilibrium for the intermediate range of magnetic field strengths (the realm of the MWDs) existed, even for hydrogen gas.
However, we recently conducted a comprehensive evaluation of hydrogen ionization balance in arbitrary magnetic fields, and we showed that field effects are significant in the conditions found in the atmospheres of the strongest magnetic white dwarfs \citep{vera2020}. Although our chemical model considers the most basic chemical species (neutral atoms, protons, free electrons) and does not include complexes such as molecules, particle chains and negative ions, this represents a step toward a comprehensive physical representation of the gas in the atmospheres of magnetic white dwarfs.

The present work is aimed at solving the inconsistency of current model atmospheres of hydrogen-rich magnetic white dwarfs (DAH stars), which use field-dependent opacities but ignore the magnetic effects on the particle abundances.  Detailed evaluations of atomic hydrogen populations in magnetic fields involve changes in the structure of atoms beyond the Zeeman perturbative approach. These changes have consequences on the gas partition function. Its determination requires taking into account finite temperatures and, hence, the thermal motion of particles, with the particularity that the motion of an atom across the magnetic field affects its internal energies \citep{pavlov1993}. This demands the use of the so-called pseudomomentum to separate the center-of-mass motion of the atom from the relative electron-proton motion \citep{gorkov:1968}.

Here, we apply a chemical model that utilizes fits of accurately evaluated energy levels of atoms at rest \citep{schimeczek2014b}. These evaluations are complemented by center-of-mass effects on the internal atomic structure, which arise from the finite proton mass and thermal motions across the magnetic field. For sufficiently large values of the pseudomomentum transverse to the magnetic field, our gas model considers the formation of the so-called decentered states, i.e., atomic states where the electronic wave function is shifted from the Coulomb center to a magnetic well \citep{burkova1976}. Furthermore, particle interaction effects on the chemical equilibrium are included through the usual occupation probability approach with a few updates (effective atomic sizes depending of the field strength). Beyond the chemical equilibrium calculation, all other constitutive physics in our MWD atmosphere models (opacity sources, radiative transfer of polarized light) follow the state-of-art theory as detailed below.

The structure of this paper is as follows. Basic approximations for modeling atmospheres of magnetic white dwarfs are presented in Section \ref{s.basic}. In Section \ref{s.ioniz}, we provide the chemical model that determines the occupation numbers of atomic states for a magnetized gas at ionization equilibrium. Section \ref{s.transfer} describes the method used to solve the radiative transfer equations for polarized light represented by the four Stokes parameters, while Section \ref{s.opacity} lists the relevant opacity sources and magnetooptical parameters used in the transfer calculation.  In Section \ref{s.surface}, we discuss the integration method of the Stokes intensities over the observed stellar hemisphere. In Section \ref{s.results}, we compare SDSS observed spectra to the theoretical spectra as a check of the new numerical code. We analyze how the improved constitutive physics used in our code affects the determination of physical parameters of strong magnetic white dwarfs (Sect. \ref{s.effects}). Additionally, predictions of our spectral energy distribution fits for a group of magnetic white dwarfs are compared with previous studies in the literature (Sect. \ref{s.previous} and Sect. \ref{s.additional}). A final discussion and conclusions follow.

\section{Basic assumptions}\label{s.basic}

The models presented here are based on pure hydrogen local thermodynamics equilibrium (LTE), plane-parallel atmospheres. 
Since there are no calculations to date that fully account for the effects of magnetic fields on the hydrostatic structure of MWDs \citep{stepien1978, landstreet1987, ferrario2020}, we assume that magnetic pressure is negligible in the outer layers of the atmosphere where the spectrum of a star originates \citep{wickramasinghe1988, jordan1992}. Therefore, we used zero-field models to compute the temperature and pressure distributions throughout the atmospheric layers \citep{rohrmann2012}. We also imposed the null-convective-flux condition according to results suggesting that convection is suppressed by magnetic fields \citep{jordan2001, tremblay2015, gentile2018}.
In addition, surface gravity is fixed at $\log g=8$, a typical value for white dwarfs, as applied in other MWD studies \citep{euchner2002, kulebi2009, amorim2023}.
This choice seems reasonable, given that mass determination from spectral line fitting is not fully reliable in the absence of an appropriate theory for combined Stark and magnetic broadening.

On the other hand, the field distribution on the stellar surface is assumed to be generated by a dipole that is centered or offset with respect to the barycenter of the star, for which we assume a spherical shape. 
Specifically, we chose Cartesian coordinates centered in the star, with the $z$-axis along the magnetic dipole and the line of sight forming an angle of $i$ on the $xz$ plane \citep[see][]{achileos1989}. Thus, the pole-on and equator-on views of the star are given by $i=0^\circ$ and $i=90^\circ$, respectively.  With distances measured in stellar radius units, $r=(x^2 + y^2 + z^2)^{1/2}=1$ defines the stellar surface. The dipole center is located at the position $\mb{a}=(a_x,a_y,a_z)$. Consequently, the magnetic field is described by
\beq\label{field}
\mb{B} = \frac{B_{d}}{2r'^5}  \left(3x' z', 
        3y' z', 3z'^{2}-r'^{2}\right), \hskip.15in
r'=(x'^2 + y'^2 + z'^2)^{1/2},
\eeq
where $w'=w-a_w$ ($w=x,y,z$) and $B_\text{d}$ equal the polar field strength at the stellar surface for the case of a centered dipole ($\mb{a}=0$). If the dipole is off-center, the field strengths at the poles are $B_\text{d}(1\pm a)^{-3}$. Through the parameters $B_\text{d}$, $i$, and $\mb{a}$, Eq. (\ref{field}) yields a nonuniform distribution over the stellar surface.
The dipole configuration is the simplest and most important term in a multipolar expansion of the magnetic potential. Such an approximation resulted to be adequate for calculating emerging flux from magnetic white dwarfs \citep{martin1984, achileos1989, kulebi2009}, especially if one takes into account that in some cases it is not easy to distinguish between an offset dipole and a combination of multipoles \citep{martin1984, putney1995}.

\section{Ionization equilibrium and occupation numbers}\label{s.ioniz}

We give special attention to the determination of occupation numbers of hydrogen atoms in a magnetic field. The details of evaluations used in the present work are described in depth by \citet{vera2020}. Here, only a short synopsis is provided.
 
The inner state of a hydrogen atom is specified by a set $\kappa$ of quantum numbers, for which one usually chooses the asymptotic ones corresponding to the Coulomb approximation at the weak-field limit,  $\kappa=\{n,l,m,m_s\}$ ($\beta\ll1$, $\beta= B/B_0$, $B_0 \approx 4.70103\times 10^9$~G), or those of the high-field Landau regime,  $\kappa=\{N,\nu,m,m_s\}$ ($\beta\gg1$), where $n$, $l$, $m,$ and $m_s$ are, respectively, the  principal, azimuthal, magnetic and spin quantum numbers, whereas $N$ is the number of the Landau level and $\nu$ is the usually called the longitudinal quantum number associated with atomic excitations in the magnetic-field direction. 
Both sets share $m$ and $m_s$, while the remaining quantum numbers are connected by mathematical relations first given in \citet{vera2020}. 

The traslational state of an atom is labeled by the eigenvalue, $\mb{k}=(k_\perp,k_z),$ of the so-called pseudomomentum operator, with $k_\perp$ and $k_z$ being the components transversal and parallel to the field, respectively. The pseudomomentum is associated with the translational invariance of the Hamiltonian of the atom in a magnetic field \citep{gorkov:1968}.
The binding energy of a magnetized atom and its motion perpendicular to the field $\mb{B}$ are generally not separable \citep{pavlov1993}. The total energy of the atom is expressed by
\beq \label{e.EE}
E = \cm{E}_\kappa(k_\perp) +\frac{k_z^2}{2m_\text{H}},
\eeq
where $m_\text{H}$ is the atom mass. Explicit analytical approximations of $\cm{E}_\kappa(k_\perp)$ are given in \citet{vera2020} for any combination of inner and translational states. These evaluations include energy data for atoms at rest \citep{schimeczek2014b} and the formation of the so-called decentered states that arise in atoms with high pseudomomentum transverse to the field \citep{potekhin2014b}.

The number densities of atoms ($n_\text{H}$), electrons ($n_\text{e}$), and protons ($n_\text{p}$) are obtained from LTE conditions for an electrically neutral hydrogen gas ($n_\text{e}=n_\text{p}$) at a given temperature, $T$. The ionization equilibrium for hydrogen in a magnetic field is given by
\beq \label{e.Q}
\frac{n_\text{H}}{n_\text{e}n_\text{p}}
=\frac{\lambda_e^3  Z_\text{H}}{2} f(\eta)
,\eeq
with
\beq \label{e.Qf}
f(\eta)= \frac{\tanh(\eta)(1-e^{-q\eta})}{q\eta^2},\hskip.3in
q = \frac {2m_\text{e}}{m_\text{p}},
\eeq
\beq \label{e.x}
\lambda_\text{e}=\hbar\sqrt{\frac{2\pi}{\kB T m_\text{e}}},\hskip.3in
\eta = \frac{\hbar\omega_\text{e}}{2\kB T},\hskip.3in
\omega_\text{e}=\frac{eB}{m_\text{e}c}
,\eeq
where $Z_\text{H}$ is the internal partition function of the atoms, $\lambda_\text{e}$ the electron thermal wavelength, $\omega_\text{e}$ the cyclotron frequency, $\hbar=h/2\pi$, $h$ the Planck constant, $\kB$ the Boltzmann constant, $c$ the speed of light, $e$ the electron charge, and $m_\text{e}$ and $m_\text{p}$ the electron and proton masses. The factor $f(\eta)$ comes from the excess of chemical potential from free electrons and ionized atoms. 
The internal partition function contains nonideal effects and the coupling of internal and transverse kinetic energies,
\beq \label{e.ZH}
Z_\text{H}= \sum_\kappa\frac{1}{m_\text{H}\kB T}  \int w_\kappa(k_\perp)
   e^{-\cm{E}_\kappa(k_\perp)/(\kB T)}k_\perp dk_\perp ,
\eeq
with $w_\kappa(k_\perp)$ being the so-called occupational probability of the state $(\kappa,k_\perp)$. The quantity $w_\kappa(k_\perp)$ represents a reduction of the phase space available in an atom due to interactions with other particles.
The number density of atoms in an inner state, as required in opacity evaluations, is determined by
\beq \label{e.nkappa}
n_\kappa = \frac{n_\text{H}}{m_\text{H}\kB T Z_\text{H}} \int w_\kappa(k_\perp)
   e^{-\cm{E}_\kappa(k_\perp)/(\kB T)}k_\perp dk_\perp .
\eeq
In the zero-field limit, $\eta\rightarrow 0$, $f(\eta)\rightarrow 1$, and the usual Saha function is recovered from Eq. (\ref{e.Q}), with
\beq \label{e.ZH}
Z_\text{H} \rightarrow \sum_\kappa w_\kappa e^{-\epsilon_\kappa/(\kB T)},
\hskip.2in
\cm{E}_\kappa(k_\perp) \rightarrow  \epsilon_\kappa +\frac{k_\perp^2}{2m_\text{H}}, 
\hskip.2in
 (\beta\rightarrow 0)
,\eeq
where $w_\kappa$ is the occupational probability and $\epsilon_\kappa$ the binding energy of the state $\kappa$, both uncoupled from the translational motion. Furthermore, in field-free conditions the pseudomomentum of an atom is reduced to the usual canonical momentum. \footnote{Actually, only the transversal component is affected by the field as a motion integral associated with the translational invariance of the atom.}

Free electron dynamics shows a continuum of energy due to movements along the field direction and quantized contributions (units of the cyclotron energy) from the perpendicular direction. In the absence of particle perturbations, it is given by 
\beq\label{continuum}
E=\hbar\omega_\text{e}\left(N+m_s+\frac12\right)+\frac{k_z^2}{2m_\text{e}}, \hskip.1in N=0,1,2,\dots, \hskip.1in m_s=\pm\frac12,
\eeq
where a giromagnetic ratio of $g_\text{e}=2$ has been assumed. These energy values are required for evaluating the ionization equilibrium and photoionization thresholds.

\section{Radiative transfer}\label{s.transfer}

Radiative transfer of polarized light in a plane-parallel atmosphere with an LTE source function is described by the following four coupled differential equations over the four Stokes parameters $\{I, Q, U, V\}$ \citep{beckers1969, hardorp1976}:
\bea \label{RTransfer}
\mu\frac{dI}{d\tau} & = & \eta_{I}\left(I-B_\text{p}\right)+\eta_{Q}Q+\eta_{V}V,\cr
&&\cr
\mu\frac{dQ}{d\tau} & = & \eta_{Q}\left(I-B_\text{p}\right)+\eta_{I}Q+\rho_{R}U,\cr
&&\cr
\mu\frac{dU}{d\tau} & = & \rho_{R}Q+\eta_{I}U-\rho_{W}V,\cr
&&\cr
\mu\frac{dV}{d\tau} & = & \eta_{V}\left(I-B_\text{p}\right)+\rho_{W}U+\eta_{V}V,
\eea
with $B_\text{p}$ being the source function assumed to be the Planck function, $\mu=\cos\theta$, $\theta$ the angle between the direction of light propagation and the normal of the stellar surface ($z'$ axis in a local coordinate system), $\tau$ the optical depth, $d\tau=-\kappa_{p}dz'$, $\kappa_{p}$ the Rosseland mean opacity evaluated with the unpolarized continuum absorption coefficient, $\rho_{R}$ and $\rho_{W}$ magneto-optical parameters, and $\eta_I$, $\eta_Q$, and $\eta_V$ combinations of the absorption coefficients given by
\bea
\eta_{I} & = & \frac{1}{2}\eta_{p}\sin^{2}\psi+\frac{1}{4}\left(\eta_{l}+\eta_{r}\right)\left(1+\cos^{2}\psi\right),\\
&&\cr
\eta_{Q} & = & \left[\frac{1}{2}\eta_{p}-\frac{1}{4}\left(\eta_{l}+\eta_{r}\right)\right]\sin^{2}\psi, \\
&&\cr
\eta_{V} & = & \frac{1}{2}\left(\eta_{r}-\eta_{l}\right)\cos\psi.
\eea
Here, $\psi$ is the angle between the direction of light propagation and the direction of the magnetic field, while $\eta_{l}$, $\eta_{p}$, and $\eta_{r}$ are the monochromatic absorption coefficients (normalized to $\kappa_{p}$) originating from atomic transitions due to radiation with left-circular ($\Delta m=-1$), linear ($\Delta m=0$), and right-circular ($\Delta m=+1$) polarizations, respectively.

Our code solves Eqs. (\ref{RTransfer}) using the semi-analytical method of \citet{martin1979} for a grid of optical depths $\{\tau_j\}$, where it is assumed that
\beq
X=X_{a}+X_{b}\tau+\sum_{i=1}^{4}X_{c,i}\exp\left(a_{i}\tau\right)
\eeq
between two successive layers ($\tau_j\leq\tau\leq\tau_{j+1}$) for the Stokes intensities $X=I,Q,U,V$, with $X_a$, $X_b$, and $X_{c,i}$ constants. The evaluation method starts from an initial condition of \citet{unno1956} at the inner boundary and then follows an iterative procedure from the deepest layer to the outermost one. Within the limit of a field-free medium ($B\rightarrow 0$), $\rho_R=\rho_W=0$, $\eta_r=\eta_l=\eta_p$, $\eta_Q=\eta_V=0$, and the usual single radiative-transfer equation on $I$ is obtained.

\section{Opacity sources}\label{s.opacity}

Line absorption cross-sections were calculated with the \textit{h2db} database of \citet{schimeczek2014}, which constitutes the most recent and complete ones for computing energies and oscillator strengths of an isolated hydrogen atom in arbitrary magnetic fields. Bound-bound cross-sections are represented by
\beq
\sigma_{\text{bb}}=\frac{\pi e^2 f}{m_\text{e} c}\frac{\Re\left[W\left(Z_l\right)\right]}{\sqrt{2\pi}\Delta_D}, 
\label{L-Broad}
\eeq
where $f$ is the oscillator strength of the transition,  $\omega_l$ the line-center frequency, and $\Delta_D$ the thermal broadening
\beq\label{Doppler}
\Delta_D = \left ( \frac{2k_B T}{m_\text{H} c^2} \right )^{1/2} \omega_l .
\eeq
In Eq. (\ref{L-Broad}), $\Re\left[W (Z_l)\right]$ represents a Voigt profile \citep{faddeyeva1961, armstrong1967}, with 
\beq\label{Faddeyeva}
W(Z_l)=e^{-Z_l^2} \left (1+\frac{2i}{\sqrt{\pi}} \int_{0}^{Z_l}e^{t^2} \text{d}t \right )
\eeq  
and
\beq\label{Zfun}
Z_l=\frac{\omega-\omega_l+\frac{1}{2}\Delta_S i}
 {\sqrt{2}\Delta_D}.
\eeq  
In the last equation, $\Delta_S$ is the Stark broadening, which was calculated according to \citet{jordan1992}:
\beq\label{Stark}
\Delta_S = 0.0192 c F_0 \overline{n_k} C,
\eeq
with $F_0$ being the Holtsmark normal field strength, $\overline{n_k}$ an average value calculated from the lower and upper $n$ quantum numbers \citep{unsold1968, rauch1991}, and $C$ a free parameter that was set equal to 0.1 \citep[see][]{putney1995}. Eq. (\ref{Stark}) is just a rough approximation, since no comprehensible data about the effects of simultaneous arbitrary magnetic and electric fields on the hydrogen atom have been published so far. 

The knowledge of the bound-free opacity of atoms at magnetic fields is still fragmentary. Its calculation demands considerable work since it involves numerous transition channels \citep{jordan1989,west1989}, many initial states, and more sophisticated wave functions than those present in the zero-field photoionization process. A number of rigorous evaluations have been performed, but they are limited in terms of the field strength domain, wavelength range, and number of bound states and/or light polarizations \citep[e.g.,][]{kara1981, bhattacharya1985, delande1991, wang1991, merani1995, zhao2007, zhao2016, zhao2021}. Therefore, we followed the usual treatment based on transition probabilities calculated at the rigidity approximation for the wave functions \citep{lamb1974}, where the allowed transitions ($\Delta l=\pm1$, $\Delta m=0,\pm1$) from bound states to Landau states are distributed according the Wigner-Eckart theorem, and the wavelengths of all bound-free edges are calculated with accurate energies of bound and free states as functions of the field strength (Eqs. (\ref{e.EE}) and (\ref{continuum})). The precision of this method has been discussed by \citet{jordan1995}.

A cyclotron absorption process is produced by $\Delta m=+1$ free-free transitions (i.e., right hand circularly polarized light)\footnote{Transitions with $\Delta m=0$ and $-1$ are forbidden by conservation laws in the kinematics \citep{lamb1972}.} of electrons from a lower Landau state into a more energetic one, with a peak near the cyclotron resonance due to transitions between adjacent levels. The corresponding cross-section is calculated as 
\beq \label{cyc} 
\sigma_\text{cy}=\frac{\Re\left[W\left(Z_c\right)\right]}
 {\sqrt{2\pi}\Delta_\text{e}}\sigma_+, 
\eeq
with
\beq
Z_c=\frac{\omega-\omega_\text{e}+\nu_\text{eff} i}
{\sqrt{2}\Delta_\text{e}}
\eeq 
and
\beq\label{Doppler_e}
\Delta_\text{e} = \left ( \frac{2k_B T}{m_\text{e} c^2} \right )^{1/2} \left | \cos \left( \psi \right ) \right | \omega_\text{e},
\eeq 
where $\sigma_+$ is the frequency-integrated cross-section calculated quantum-mecanically as \citep{lamb1974}
\beq
\sigma_+ \approx \frac{e^2}{\hbar c} \left (\frac{2\pi c}{\omega_\text{e}} \right )^2 \left (\frac{B e \hbar}{m_e^2c^3} \right ) \frac{\omega_\text{e}}{1-e^{-\hbar \omega_\text{e} / k_B T}},
\label{CSection}
\eeq
and $\nu_\text{eff}$ is half the half-width of a Lorentzian profile describing collisions by electrons \citep{bekefi1966}. 
\begin{figure}
\includegraphics[width=.48\textwidth]{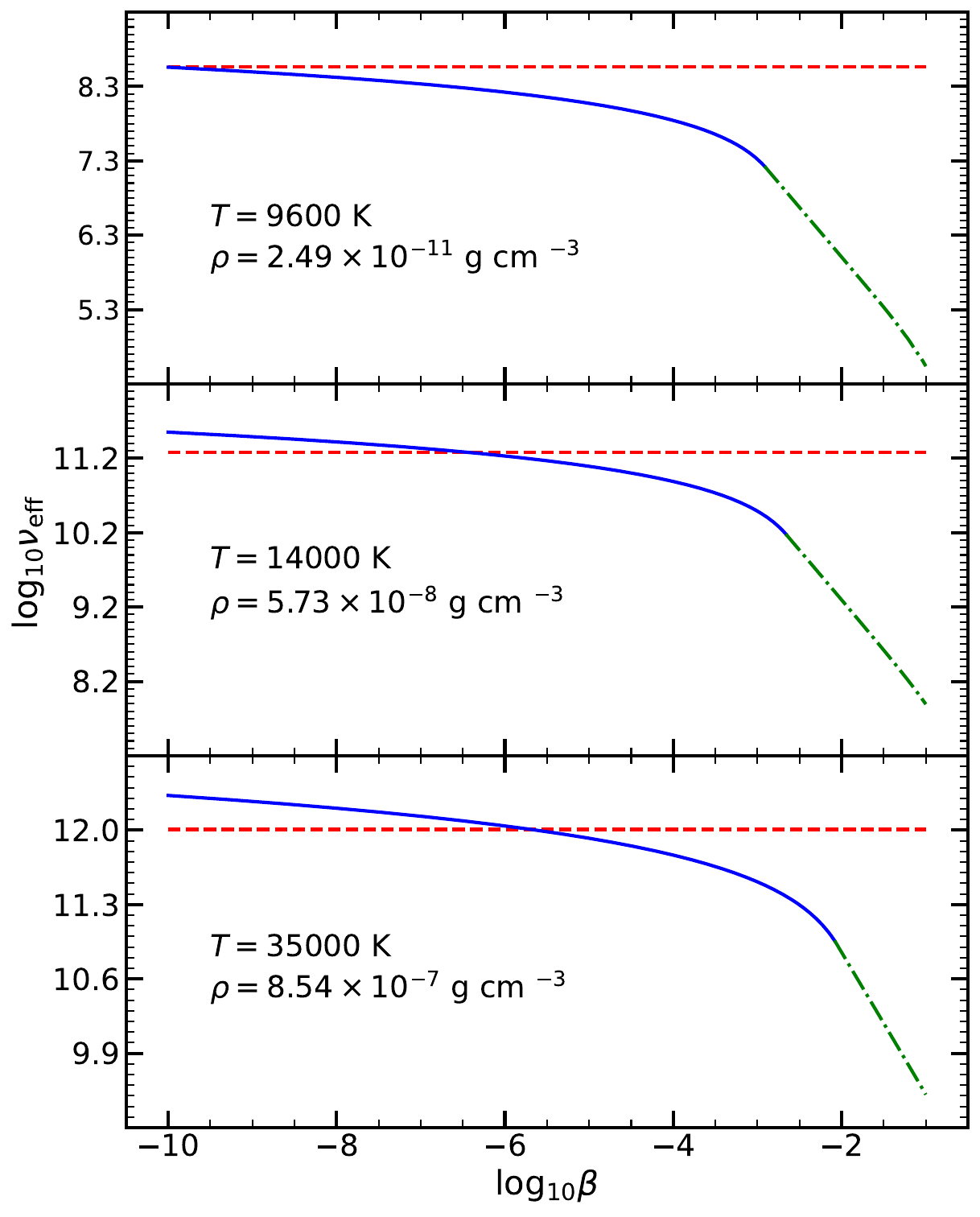}
\caption{Dependence of frequency of electron-electron collisions ($\nu_\text{eff}$) with field strength as predicted by Eq. (\ref{collision1}) (continuum line for $B < 6.3T^{3/2}$, dash-dotted line for $B > 6.3T^{3/2}$) and Eq. (\ref{collision2}) (dashed line). The selected physical conditions (gas temperature $T$, mass density $\rho$) correspond to different depths in a model atmosphere with $T_\text{eff}=15000$~K.} 
\label{veff}
\end{figure}   
This is given by \citep{zheleznyakov1999}
\beq\label{collision1}
\nu_\text{eff}= \left\{ \begin{array}{ll}\displaystyle
 \frac{2\sqrt{2\pi}}{5\exp(1)}
 \frac{m_e^{1/6}c^{4/3} (\kB T)^{1/2}n_{p}}{B^{4/3}}, 
& B > 6.3T^{3/2},\\
[3ex]\displaystyle 
 \frac{8\sqrt{2\pi}}{15}\frac{e^{4}n_{p}}{m_e^{1/2}\left(\kB T\right)^{3/2}}
\ln \left(\frac{m_e^{1/2}c (\kB T)^{3/2}}{e^{3}B}\right), & 
  B\le 6.3T^{3/2}.
\end{array} 
\right.
\eeq
For low enough magnetic fields, Eq. (\ref{collision1}) gives unphysical values (logarithmic divergence). To avoid that, we adopted the zero-field expression of \citet{ginzburg1967} as the upper limit for the $\nu_\text{eff}$ value (see Fig. \ref{veff}):
\beq \label{collision2}
\nu_\text{eff}=\frac{2\sqrt{2\pi}}{3}\frac{e^4n_p}{m_e^{1/2}\left(\kB T\right)^{3/2}}\ln\left(\frac{k^3T^3}{4\pi n_p e^6}\right).
\eeq   

Finally, the code includes the calculation of magneto-optical parameters, Faraday rotation ($\rho_R$), and Voigt effect ($\rho_W$), which describe anomalous dispersion of light. While $\rho_R$ arises from the inequality between refractive indexes for right and left circularly polarized light in the presence of a magnetic field, $\rho_W$ comes from a phase shift between linear polarized components of the electric-field vector, both parallel and perpendicular to the magnetic field. Magneto-optical parameters have contributions from both lines and continuum, as described by \citet{martin1981, martin1982}, and \citet{jordan1991}. 
Specifically, line contributions are given by
\beq
\rho_R=-\left(\sum_i \eta_{ri}F_{ri}-\sum_j\eta_{lj}F_{lj}\right)\cos\psi,
\label{FarRot}
\eeq
\beq
\rho_W=-\left(\sum_i \eta_{ri}F_{pi}-\frac{1}{2}\sum_j\eta_{lj}F_{lj}\frac{1}{2}\sum_k\eta_{rk}F_{rk}\right)\sin^2\psi,
\label{VoiEff}
\eeq
with 
\beq
F=\frac{1}{a}\left(\frac{1}{2}xV+\frac{1}{4}\frac{\partial V}{\partial x}\right)
\eeq
being the dispersion function \citep{wittmann1974, martin1981}, where $a=\Im\left(Z_l\right)$, $x=\Re\left(Z_l\right)$, $V=\Re\left[W\left(Z_l\right)\right]$, and
\beq
\frac{\partial V}{\partial x}=-\frac{\Re\left[Z_l\cdot W \left(Z_l\right)\right]}{\sqrt{\pi}\Delta_D^2},
\eeq
$W$ and $Z_l$ being given by Eqs. (\ref{Faddeyeva}) and (\ref{Zfun}), respectively. In Eqs. (\ref{FarRot}) and (\ref{VoiEff}), the sums extend over all allowed transitions. On the other hand, continuum contributions to $\rho_R$ and $\rho_W$ were computed as they were by \citet{kulebi2010}, which featured a self-consistent calculation from kinetic theory of plasmas to determine refractive and absorptive properties of magnetic atmospheres, such that
\beq
\rho_R=-\frac{\sqrt{\pi}}{2c}\frac{\omega_p^2}{\sqrt{2}\Delta_{D}}\Im\left[W\left(Z_c\right)\right]\cos\psi,
\eeq
\beq
\rho_W=-\frac{\sqrt{\pi}}{4c}\frac{\omega_p^2}{\sqrt{2}\Delta_D}\Im\left[W\left(Z_c\right)\right]\sin^2\psi,
\eeq
where $\omega_p=\sqrt{4\pi n_e e^2/m_e}$ is the plasma frequency.

\begin{figure}
\begin{centering}
\includegraphics[width=.48\textwidth]{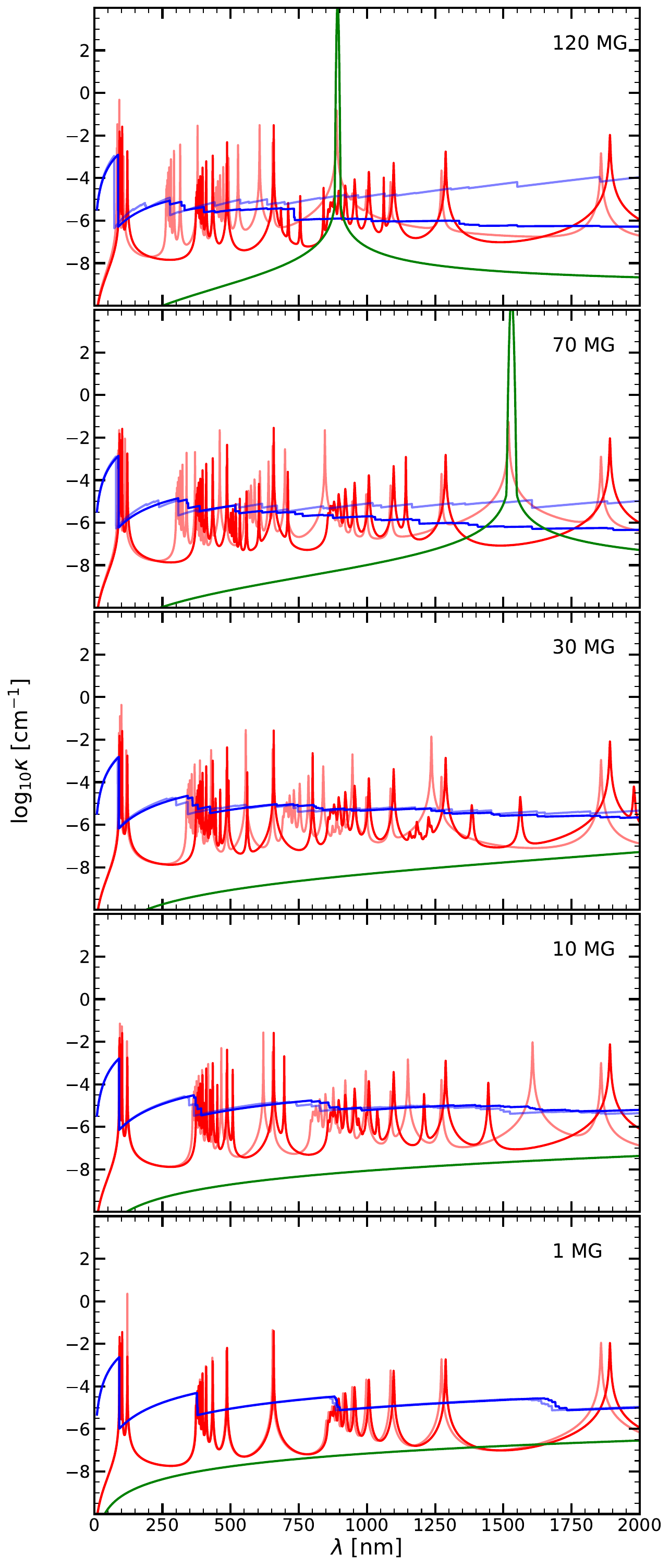}
\par\end{centering}
\caption{Absorption coefficients for bound-bound (red), bound-free (blue; dark and light lines for $\Delta m=1,-1$ transitions), and cyclotron (green) processes for various magnetic field strengths.} 
\label{Opacities}
\end{figure}   
%
In Fig. \ref{Opacities}, we compare the different contributions of hydrogen opacity arising from transitions $\Delta m=\pm 1$ in a gas with $T=20000$~K for different magnetic field strengths. For low enough magnetic strengths ($B\la1$~MG), opacities from different light polarizations tend to be coincidental and resemble those of a field-free gas. As magnetic field strength increases, photoionization continua and spectral lines split in a number of components.
Besides this, cyclotron absorption is dominant around cyclotron wavelength and becomes negligible far away it, with the peak moving from far- to near-infrared wavelengths as the field strength increases.

\section{Surface integration method}\label{s.surface}

Equation (\ref{field}) provides the magnetic-field value in each point of the stellar atmosphere affecting particle populations, opacities and radiative energy transfer. Because of the high surface gravity in white dwarfs, their atmospheres are compact (thickness much lower than the stellar radius), and the depth dependence of the magnetic field as given by Eq. (\ref{field}) can be neglected. Individual model atmospheres are computed for a selection of $M$ points centered in cells that cover the visible hemisphere of the star. Transfer equations (\ref{RTransfer}) are solved in these points, and the resulting Stokes intensities are integrated with a quadrature scheme to obtain the emerging stellar flux,
\beq\label{average}
X=\sum_{j=1}^{M}w_j X_{j}, 
\eeq
where $w_j$ is a quadrature weight and $X_{j}=I_{j}$, $Q_{j}$, $U_{j}$, $V_{j}$ are the emerging intensities in the $j$th-cell.

%
\begin{figure}
\includegraphics[width=.49\textwidth]{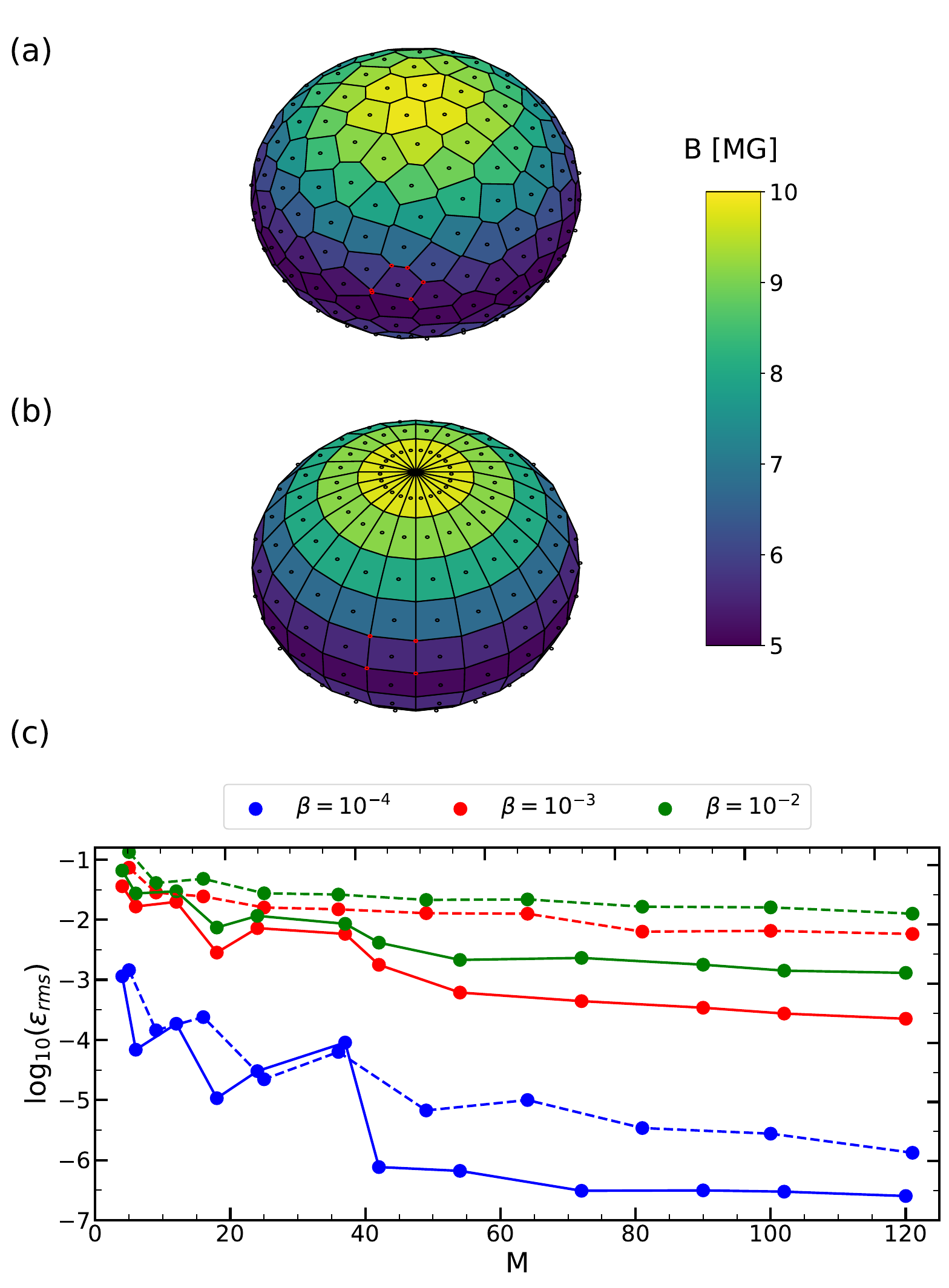}
\caption{Segmentations of stellar surface and their performances in energy-flux integrations for a centered dipole ($B_{d}=10$~MG, $i=45^{\circ}$). The magnetic-field strength in each cell is averaged using its values on the center (black circle) and vertices (red circles). 
{\it Upper panel}: Segmentation based on spherical $t$-design ($t=19$) and its associated Voronoi tessellation. 
{\it Middle panel}: Segmentation constructed with equal steps in latitude and longitude. 
{\it Lower panel}: Variation of mean squared error of emerging flux as a function of the cell number for $t$-design (solid lines) and constant latitude and longitude step (dashed lines) integrations and different field strengths.} 
\label{Quads}
\end{figure}   
%
%
\begin{table*} 
\caption{Stellar parameters of a selection of MWDs derived from SDSS spectra fits in the present work and other recent studies.
}  \label{table1}
\begin{tabular}{c|crrr|rrrr|rrrr}
\hline 
\hline
 & & This work & & & & Hardy+ & & & &  Amorim+ & & \\
\hline
 Star  & $T_\text{eff}$[K] & $B_\text{d}$[MG] & $i$[$^\circ$] & $a_{z}$  &
         $T_\text{eff}$[K] & $B_\text{d}$[MG] & $i$[$^\circ$] & $a_{z}$  &
         $T_\text{eff}$[K] & $B_\text{d}$[MG] & $i$[$^\circ$] & $a_{z}$  \\
\hline 
J0725+3214 & 26000  &  13.16 &  80    & $-0.21$  &
             22240  &  13.82 &  52    &  0.06    &
             24000  &  15.15 &  58    &  $0.29$ \\
J0805+2153 & 39000  &   7.00 &  85    & $-0.15$  &
             30000  &   3.00 &  17    &  0.30    & 
             37141  &   6.77 &  66    & $-0.49$ \\
J0931+3219 & 11000  &   8.46 &  75    &  $0.19$  &
             13476  &   9.04 &  52    &   0.30   & 
             18000  &  12.95 &  15    &  $0.26$ \\
J1018+0111 & 10500  & 108.12 &  60    &  $0.1^*$ &
             12845  &  76.07 &  16    &  0.05    &
             11000  & 122.69 &  65    &  0.01   \\
J1511+4220 & 12000  &  11.28 &  40    &  $0.28$  &
             11595  &  14.01 &  40    &  0.30    & 
             11500  &  12.83 &  35    & $-0.28$ \\
J1516+2746 & 33000  &   3.25 &   5    &  $0.25$  &
             --     &  --    &  --    &   --     &
             30000  &  3.03  &  75    & $-0.44$ \\
J1603+1409 & 10000  &  47.01 &  70    &  $0.15$  &
             10547  &  48.88 &  47    &  0.23    & 
              9500  &  49.40 &  52    & $-0.24$ \\
J2247+1456 & 19000  & 437.10 &  10    & $-0.15$  &
              --    &  --    &  --    &   --     &
              18000 & 515.09 &  29    &  0.21   \\
\hline 
\end{tabular}
\tablefoot{(*)  Additional $a_y=0.07$ dipole offset.}
\end{table*}
%
%
\begin{table} 
\caption{For the studied MWDs, S/N of spectra, and percentage error $\chi^2$ derived from our fits using centered and offset dipole configurations.
}  \label{table2}
\begin{tabular}{crrr}
\hline 
\hline
 Star  &  S/N$\hskip.05in$  & centered  &  offset   \\
\hline 
J072540.8+321401.1 & 07.43 & 1.39$\hskip.13in$ &  1.32$\hskip.05in$ \\ 
J080502.3+215320.5 & 18.52 & 1.99$\hskip.13in$ &  1.96$\hskip.05in$ \\
J093126.1+321946.1 & 07.49 & 1.41$\hskip.13in$ &  1.29$\hskip.05in$ \\
J101805.0+011123.5 & 49.54 & 52.11$\hskip.13in$ & 40.42$\hskip.05in$ \\
J151130.2+422023.0 & 19.79 & 4.97$\hskip.13in$ &  3.88$\hskip.05in$ \\
J151606.3+274647.0 & 14.23 & 2.33$\hskip.13in$ &  2.10$\hskip.05in$ \\
J160357.9+140930.0 & 16.91 & 6.74$\hskip.13in$ &  5.00$\hskip.05in$ \\
J224741.46+145638  & 29.37 & 36.12$\hskip.13in$ & 29.23$\hskip.05in$ \\
\hline 
\end{tabular}
\end{table}

We used a quadrature integration based on spherical $t$-designs. A spherical $t$-design is a set of $N$ points distributed on the stellar surface ($\mathbb{S}^2$) for which the average value over these points of any spherical polynomial with a degree of at most $t$,  $p(\mb{r})$, is equal to the average value of the polynomial over the sphere \citep{delsarte1977}:\footnote{A spherical polynomial is a combination of the standing waves on the surface $\mathbb{S}$.} 
\beq
\frac 1N \sum^N_{j=1} p(\mb{r}_j)=\frac{1}{|\mathbb{S}^2|}\int_{\mathbb{S}^2} p(\mb{r})d(\mb{r}).
\eeq
In practice, we adopted $N$-point arrays as provided by \citet{hardin1996} for specific $t$ values in the $t$-design scheme. The $N$ points are uniformly distributed on the stellar surface, and over them a Voronoi spherical segmentation is calculated (upper panel in Fig. \ref{Quads}). 
The magnetic-field distribution in the stellar disk is determined for a number $M$ ($<N$) of Voronoi cells (all those that are in the visible hemisphere), where $\mb{B}$ is averaged taking into account its values at the center and vertices of each cell. The set of points in the $t$-design and the values on the magnetic field on them are finally used in Eq. (\ref{average}).

The example shown in Fig. \ref{Quads} (upper panel) corresponds to a $t$-design segmentation for $M=120$ points ($t=21$) on the visible hemisphere of a star with a centered dipolar field ($B_\text{d}=10$~MG, $i=45^{\circ}$).
For the same conditions, the middle panel in the figure shows the segmentation resulting when constant steps in latitude and longitude are taken. 
This segmentation produces a higher density of points toward the poles, while the $t$-design gives a uniform distribution over the whole stellar disk. The bottom panel in Fig. \ref{Quads} shows the accuracy in evaluating emergent energy flux for both surface mappings as a function of the point number $M$ on the visible disk. There, $\epsilon_\text{ms}$ represents the mean squared error in the convergence of the emerging flux in the $380$~nm~$\leq\lambda\leq800$~nm range (in practice, we used a reference model with $M=1000$). These evaluations correspond to  MWD models with $T_\text{eff}=20000$~K and three magnetic dipole intensities $B_\text{d}/B_0=10^{-4}$, $10^{-3}$, and $10^{-2}$. Convergence decreases as the field strength grows, but it clearly demands fewer iterations when the $t$-design is used.
Reliable results of emerging fluxes and spectral fits are obtained with  $t$-design quadratures for a reasonable cell number (typically from $M\approx 40$ for $\beta\approx 10^{-4}$ to  $M\approx 120$ for $\beta\approx 10^{-2}$) and CPU time. 
Appealing to its fast convergence, we can omit the ``magnetic broadening'' sometimes used to solve the finite discretization of the stellar atmosphere \citep{jordan1991, kulebi2009}.

\section{Model atmospheres and spectral fits}\label{s.results}

To check the new magnetic atmosphere code with improved constitutive physics, we analyzed a sample of magnetic white dwarfs with observed spectra in the SDSS survey and compared our best spectrum fits with the predictions of preexisting models. 
The procedure to find the best model for each object is by visual comparison of the observed spectral energy distribution with predictions of a set of model atmosphere calculations. The free parameters ($T_\text{eff}$, $B_\text{d}$, $i$, $a_z$) are selected through an error-reduction process using a least-squares method on the difference between observed ($f^\text{obs}_l$) and synthetic ($f^\text{cal}_l$) fluxes,
\beq
\chi^2=\frac1N \sum_l
  \frac{\left(f^\text{obs}_l-f^\text{cal}_l\right)^2}{\sigma_l^2},
\eeq
with $N$ being the flux-data number and $\sigma_l$ the data precision.

The selected sample of MWDs comprises eight stars with SDSS energy distribution from 380nm to 800nm. Since our main goal is to test the performance of the new code, the objects were chosen on the basis of their variety of field strength and effective temperature.
Table \ref{table1} summarizes the results of the analyses of the MWDs and provides a comparison with other recent theoretical studies.
The first row in the table gives the object names ordered by their right ascension, adopting the usual compact notation for SLOAN targets based on epoch J2000.0 coordinates (for the sake of clarity, full SDSS identifications are given in Table \ref{table2}). 
Further rows specify the best-fit stellar parameters calculated in the present work and those obtained in recent studies of \citet{hardy2023} and \citet{amorim2023}.
The goodness of each fit is represented by the $\chi^2$ value (Table \ref{table2}). Small errors are obtained for MWDs with weak or moderated field strength; however, as usual, fit deviations increase for high field objects where radiative transfer codes for magnetized white dwarf atmospheres have greater difficulty reproducing the observed spectra \citep[e.g.,][]{euchner2006}. Incidentally, the high-field stars in our sample (J2247+1456 and J1018+0111) have SDSS spectra with high signal-to-noise ratios (Table \ref{table2}), which exacerbates their $\chi^2$ values.

\begin{figure}
\includegraphics[width=.50\textwidth]{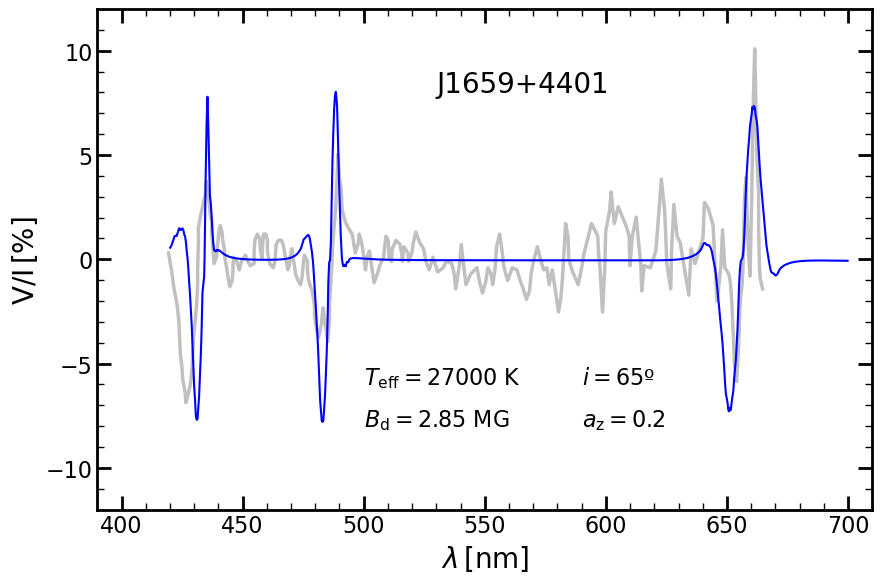}
\caption{Observed wavelength dependence of circular polarization $V$ \citep[gray line,][]{liebert1983} showing Balmer line polarization for a low field MWD (J1659+4401, also known as PG 1658+441). The blue line represents our fit model with parameter values indicated on the plot.} 
\label{Polarization}
\end{figure}

The present study focuses on the normal total intensity (Stokes parameter $I$). Although the code can also provide light-polarization information (Fig. \ref{Polarization}), this is not examined here since the data available in the literature are limited and because opacity theory becomes inappropriate, especially for evaluating the continuum polarization at high field strengths \citep[e.g.,][]{putney1995}; hence, current analyses are based in part on rather empirical relationships \citep{bagnulo2020, berdyugin2022}.

\subsection{The effect of magnetic field on populations}\label{s.effects}

Using the chemical model described in Sect. \ref{s.ioniz}, we studied the impact of the magnetic field on the populations of electronic states and analyzed how it affects the synthetic spectra of a magnetic white dwarf. 
For this purpose, we chose the SLOAN spectrum of J2247+1456, a magnetic white dwarf with the highest field of the eight stars analyzed in this paper.

\subsubsection{J2247+1456} 

The magnetic field of this object was identified by \citet{harris2003} and the average strength of its surface field was evaluated as $300$~MG. The SDSS spectrum shows features often found in strongly magnetic white dwarfs, and despite the efforts made, models cannot reproduce it very well. The observed flux was described by \citet{schmidt2003} as originating from an atmosphere of $T_\text{eff}=18000$~K and a centered dipole of 560 MG. \citet{kulebi2009} used 421.15 MG and 469.52 MG centered and offset dipoles, respectively, to fit the spectrum with an effective temperature of 50000 K. \citet{kepler2013} estimated a magnetic strength of 47.0~MG based on measurements of H$\alpha$ splitting (although this visual field determination method is inaccurate at $B\ga 100$~MG), whereas \citet{amorim2023} fit the spectrum with an offset dipole of 515.09 MG and $T_\text{eff}=18000$~K.
These studies show a range of varying results, most likely because the input physics at very strong fields is not completely clear to date.

\begin{figure}
\includegraphics[width=.50\textwidth]{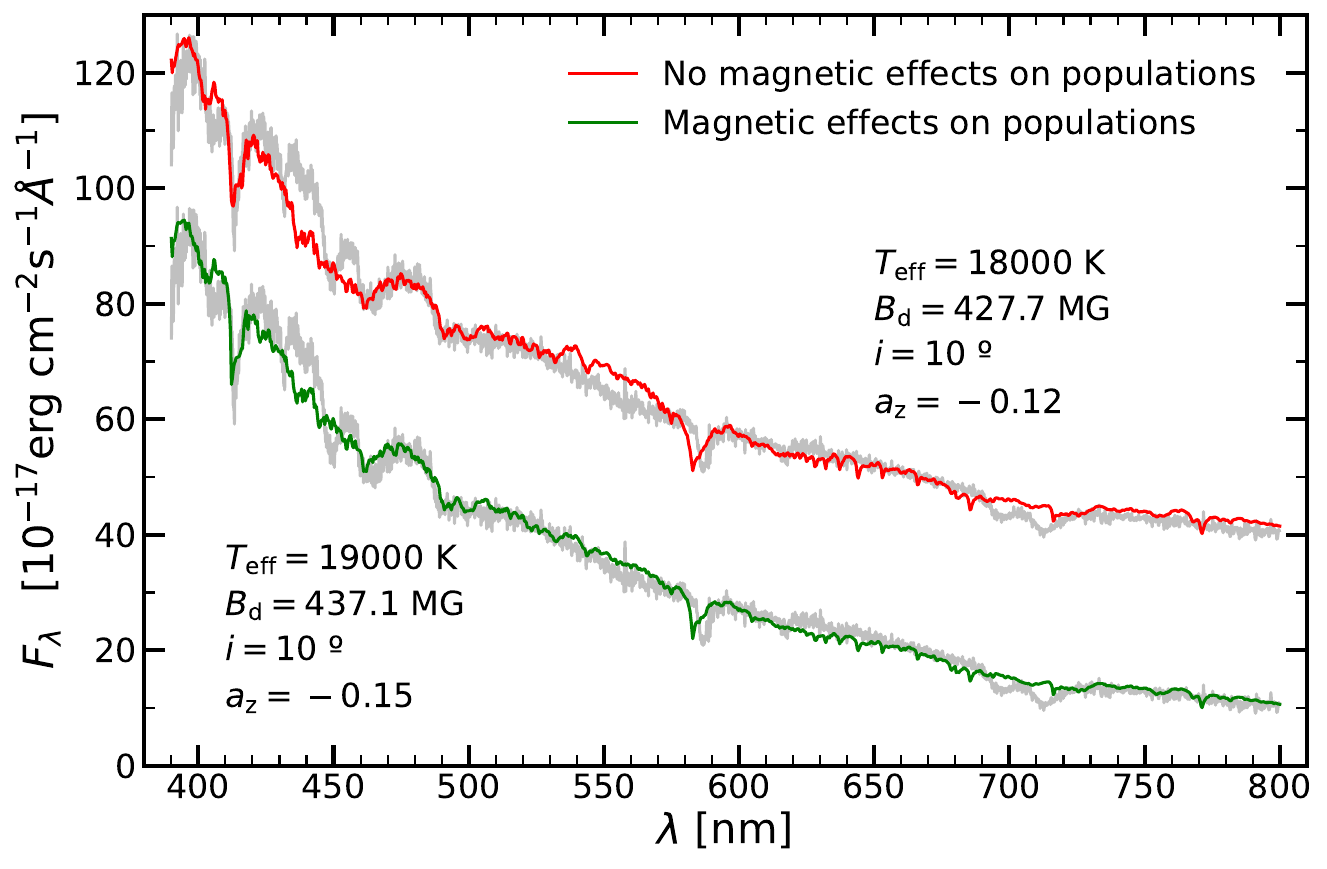}
\caption{Theoretical fluxes fit to SDSS J2247+1456 spectrum (gray lines) comparing the effects of the magnetic field on particle populations (specified on the plot). 
Upper spectrum is displaced vertically for clarity.} 
\label{PopSpectra}
\end{figure}   
In Fig. \ref{PopSpectra}, we compare the synthetic spectrum computed using our magnetic chemical model (green line) to the spectrum obtained with a standard ionization equilibrium without magnetic effect on populations (red line).
From these fits, we find field strengths of $B_\text{d}=437$~MG and 428~MG, respectively, which are relatively close to values determined from other models. On the other hand, our evaluations of the effective temperature (19000~K and 18000~K, respectively) fall near to the mean value of previous works, except for the high value obtained by \citet{kulebi2009}. 
As can be appreciated in Fig. \ref{PopSpectra}, we obtain a lightly lower spectrum fit error when magnetic effects are included in the chemical model ($\chi^2\approx 29\%$ compared to 45\% in the nonmagnetic case).

\begin{figure}
\includegraphics[width=.49\textwidth]{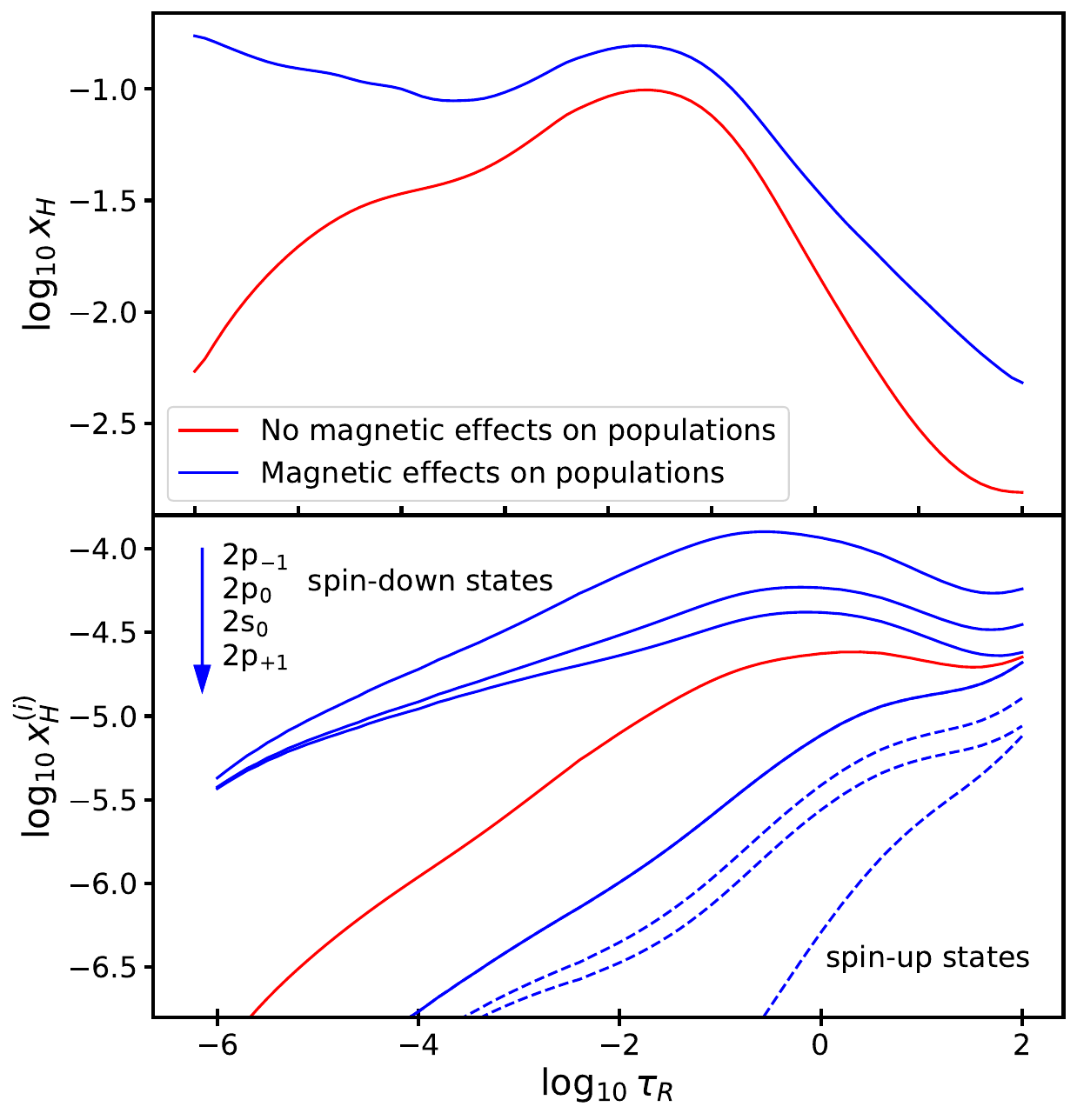}
\caption{Atomic hydrogen abundances as function of Rosseland mean optical depth for an atmosphere with $T_\text{eff}=19000$~K and  $B=437$~MG. {\it Upper panel}: Fraction of neutral atoms calculated with (blue line) and without (red line) magnetic-field effects on the ionization equilibrium. {\it Lower panel}: Abundance of atoms in $n=2$ state in absence of magnetic field (red line) and disaggregated into sublevels when the field is present (solid blue lines for spin-down and dashed blue lines for spin-up states). Abundances of $2p_{-1}(m_s=+\frac12)$ and $2p_{+1}(m_s=-\frac12)$ states are equal (lowest solid line). } 
\label{Populations}
\end{figure}

The most important result of our fits is that the inclusion of magnetic effects on the ionization equilibrium leads to a higher temperature distribution in the atmosphere of a highly magnetic white dwarf ($T_\text{eff}=19000$~K compared to $18000$~K in the nonmagnetic model).
This can be understood by analyzing the particle abundances in both models.
Effects of magnetic field on occupation numbers are clearly seen in Fig. \ref{Populations} for a model atmosphere with $T_\text{eff}=19000$~K and $B=437$~MG. 
%
%
Fig. \ref{Populations} shows the fraction of neutral atoms (upper panel) and those in sublevels of the $n=2$ state (lower panel) using our ionization model (blue lines) and a standard field-free calculation (red lines).
The first thing to note is that the magnetic field causes a significant rise in the population of neutral atoms. This is mostly due to the binding energy of the ground state $(n,l,m,m_s)=(1,0,0,-\frac{1}{2})$ increasing monotonically with the field strength, which favors the recombination process. The greatest deviations with respect to the zero-field evaluation occur in outer layers of the atmosphere where strong spectral lines are formed. Each $(n,l)$ level splits into sublevels characterized by their magnetic ($m$) and spin ($m_s$) quantum numbers. Energies of spin-up states and those with positive $m$ approach the continuum, while the spin-down states with nonpositive $m$ become tighter for conditions typical of a magnetic white dwarf atmosphere \citep{vera2020}. Consequently, sublevel populations are significantly different than those assuming a field-free gas, as can be seen in Fig. \ref{Populations} for atoms in $n=2$. 

The discussed changes in the ionization equilibrium affect the opacity and thus the emerging energy distribution from the star.
Specifically, the increase in the abundance of neutral atoms in model atmospheres that include magnetic-field effects on the particle distributions is compensated by a reduction of the gas temperature in models with abundances evaluated in the zero field. In conclusion, the use of a field-free chemical model leads to a significant underestimation of the effective temperature (a thousand degrees in the case analyzed here) for highly magnetic white dwarfs.
This discrepancy is expected to decrease for objects with weaker fields. 
However, we do not repeat a similar analysis of this effect in low field stars due to the inherent difficulties in finding optimal fits in MWD spectrum modeling. Due to the proximity of minimum values of $\chi^2$ for different combinations of free parameters \citep[which is well known; e.g.,][]{martin1984, euchner2006, kulebi2009}, field effects on particle abundances in weak magnetic stars become hidden by the search method of the best spectrum fit.

\begin{figure}
\includegraphics[width=.5\textwidth]{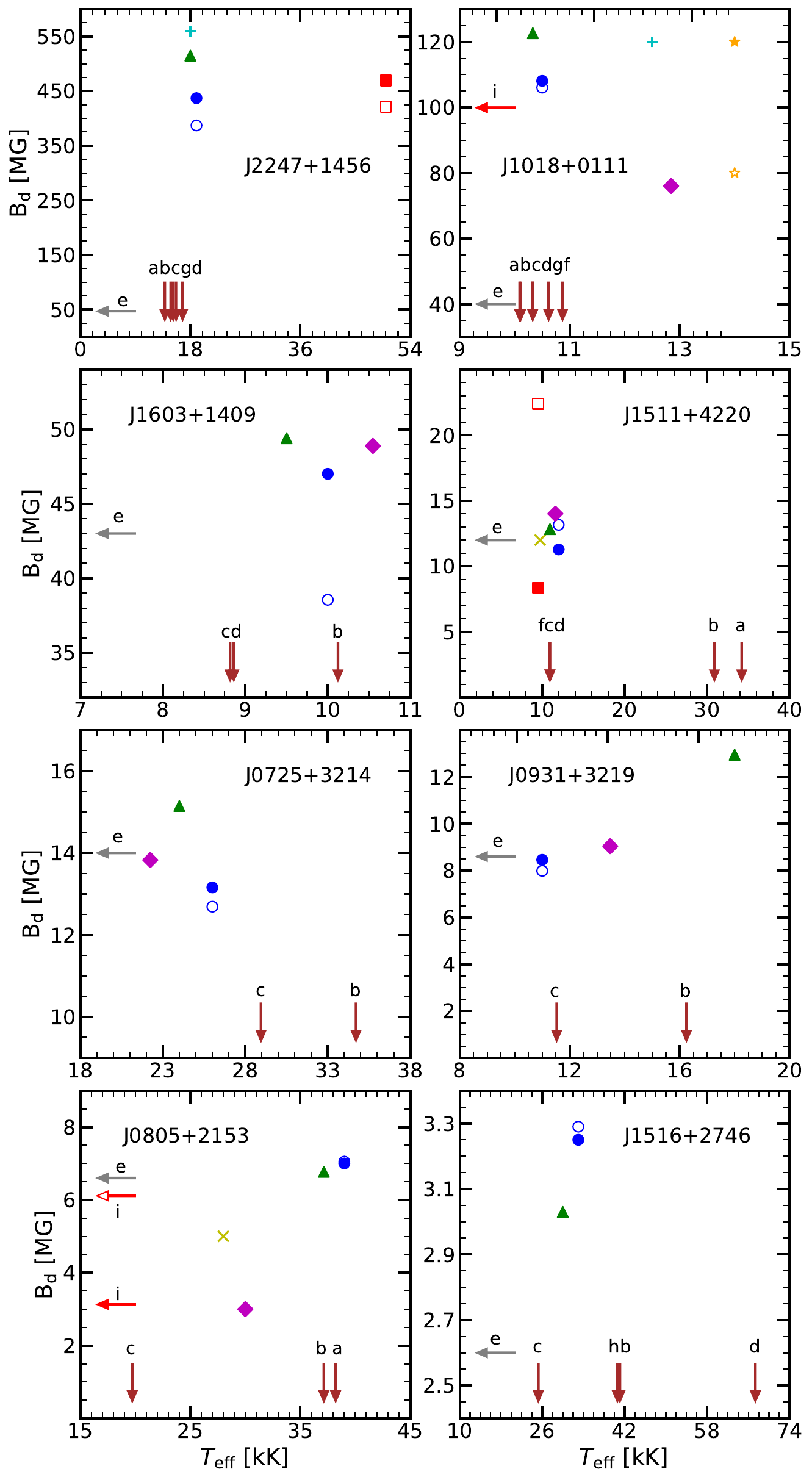}
\caption{Location of a group of MWDs in $B$--$T_\text{eff}$ diagrams, as determined from spectrum fits (symbols) and spectroscopic analysis (arrows). Vertical (horizontal) arrows mark effective temperature (field strength) obtained from spectroscopic analysis. Open (filled) symbol for centered (off-center) dipole model. Notation: a \citep{eisenstein2006}, b \citep{kleinman2013}, c \citep{dufour2017}, d  \citep{gentile2021}, e \citep{kepler2013}, f \citep{kilic2020}, g \citep{leggett2018}, h \citep{bedard2020}, i/squares \citep{kulebi2009}, diamonds \citep{hardy2023}, triangles \citep{amorim2023}, x \citep{vanlandingham2005}, crosses \citep{schmidt2003}, stars \citep{wickramasinghe1988b}, circles (this work).} 
\label{TBfits}
\end{figure}   

\subsection{Comparison with other fit studies}\label{s.previous}

Figure \ref{TBfits} shows calculated values of dipole intensity and effective temperature for the MWDs in Table \ref{table1}, which were obtained from our best spectral fit and by estimations of previous studies. Objects are ordered top to bottom and left to right by decreasing the field intensity.
Values ($B_\text{d},T_\text{eff}$) obtained from spectrum fits with magnetic models are represented by symbols, where circles correspond to results obtained in the present work.
Vertical arrows indicate $T_\text{eff}$ values derived in surveys with atmosphere models for nonmagnetic DA white dwarfs. 
Horizontal arrows point out field-strength values resulting from Zeeman splitting measurements and those obtained from flux fits with magnetic models without $T_\text{eff}$ values reported. A brief description of each star is summarized below, except J2247+1456, which was discussed in the previous subsection. 

To assess the accuracy of the new magnetic atmosphere code, below we compare our calculated spectrum for J1018+0111, J1511+4220, and J0805+2153 to those published by \citet{kulebi2009}, in both centered and offset dipole configurations. Later, in Section \ref{s.additional}, we show results from flux fits of the remaining MWDs in Table 1 
for which no published synthetic spectra were found for a direct comparison, but whose physical parameters have been derived in other works.

%
\begin{figure}
\includegraphics[width=.49\textwidth]{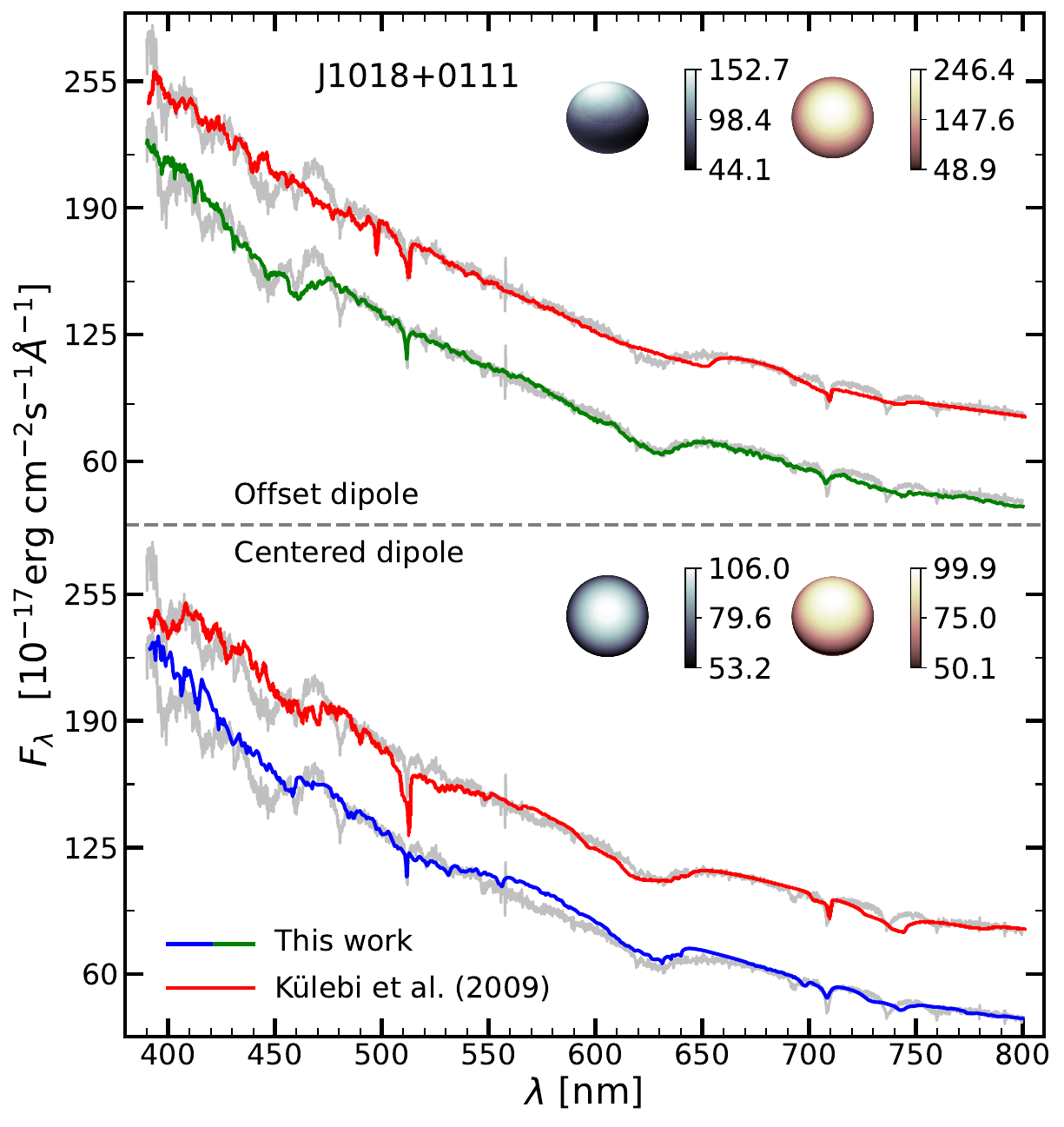}
\caption{Observed spectrum of J1018+0111 (gray lines) compared to best fits obtained in this work (green and blue lines) and by \citet{kulebi2009} (red lines), for both centered and offset dipoles (their intensity maps are shown in the inserted plots). } 
\label{j1018}
\end{figure}   
%
\subsubsection{J1018+0111} 

Also known as PG 1015+014, the first data concerning this object were published in the Palomar-Green survey \citep{green1986}. However, a previous report by \citet{angel1978} gave some details about preliminary unpublished observations of this star by the Steward Observatory and Kitt Peak National Observatory groups, who observed variable circular polarization with an amplitude of $1.5\%$ and a period of $98.75$ min, and undefined absorption features in the spectrum. These two facts were attributed to the magnetic nature of a rotating white dwarf with a surface magnetic field stronger than $100$ MG.

\citet{wickramasinghe1988b} tested different magnetic-field geometries to fit the star spectrum using a model with  $T_{\text{eff}}=14000$ K and concluded that a centered dipole is the most suitable configuration for the magnetic field, with a strength of $B_\text{d}=120$ MG and inclination of $i=70^\circ$--$110^\circ$ depending on the variability phase.  However, some spectral features could not be fit, which suggested a more complex field structure.
\citet{schmidt2003} used a centered dipole approximation to fit J1018+0111's spectrum from SDSS DR1 with $B_\text{d}=120$~MG and $T_\text{eff}=12500$~K.
A detailed study of \citet{euchner2006} using the superposition of various multipoles identified contributions to Zeeman features with fields in the range of 50--90~MG and an effective temperature of 10000~K, but it was not enough to appropriately fit to all phases. Subsequent works based on offset-dipole fits \citep{kulebi2009, amorim2023, hardy2023} found $B_\text{d}$ values in the same range as previous studies  ($\approx 76$--$123$~MG). Our best fit, $B_\text{d}=108$~MG and $T_\text{eff}=10500$~K (full circle in Fig. \ref{TBfits}), falls within that range. 

Figure \ref{j1018} compares our best fits for centered (blue line) and offset dipole (green line) models with the observed SDSS flux (gray lines) and synthetic ones of \citet{kulebi2009} (red lines). 
Our centered dipole model fails to reproduce the flux around the $\lambda 555$ region and overestimates it for wavelengths smaller than $\lambda 480$, while the fit of K\"ulebi et al. underestimates the $\lambda 515$ and $\lambda 740$ surroundings and shows other discrepancies in the bluer region. The use of an offset dipole improves both models, but some discrepancies remain, especially below $\lambda 470$ and for features redward of $\lambda 720$. However, most differences between the observed flux and our synthetic spectrum are removed in the intermediate region.

As other authors have pointed out, mentioned difficulties in achieving a good fit of the observed spectrum may be due to an oversimplification of the dipole model to describe the field distribution on the surface of J1018+0111. However, it should be taken into account that an excessive number of free parameters for the magnetic field representation can mask limitations in the approximations used to evaluate the emitted energy in each grid point of the stellar surface.

%
\begin{figure}
\includegraphics[width=.49\textwidth]{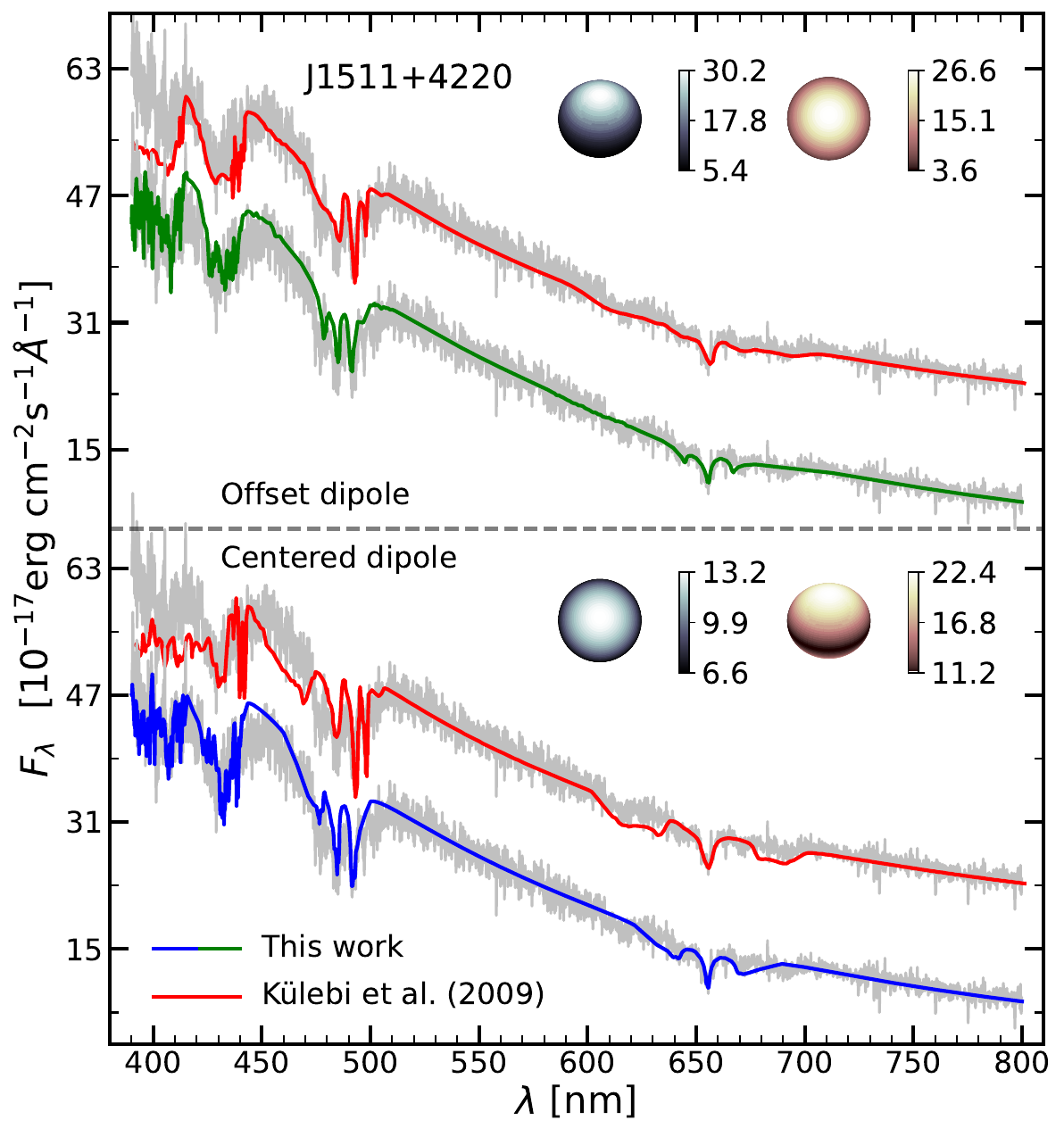}
\caption{Same as Fig. \ref{j1018}, but for J1511+4220. } 
\label{j1511}
\end{figure}   
%
\subsubsection{J1511+4220}  

This was first identified as a MWD by \citet{vanlandingham2005};  the authors derived a centered dipolar magnetic field with $B_\text{d}=12$~MG and $i=60^\circ$ through a geometric approach and an effective temperature of $9750$~K from color-color diagrams. This object was included in various studies  based on zero-field models with temperature estimations into two range: one  close 11000~K and other around 32500~K (Fig. \ref{TBfits}). 
Best spectrum fits with dipole field geometry \citep{kulebi2009, kepler2013,  amorim2023, hardy2023} were reached with $T_\text{eff}$ at the first domain and field strength in the range of $\approx 10$--$23$~MG. Our best evaluations,  $B_\text{d}=11.3$~MG and $T_\text{eff}=12000$~K, approach the lowest values previously reported.

\citet{hardy2023} categorized the SDSS spectrum of this object among those not well reproduced with a simple dipole. Fig. \ref{j1511} presents our best fits and those of \citet{kulebi2009}. Centered dipole models are not really satisfactory, although our fit (with lower field strength and slightly higher temperature, Fig. \ref{TBfits}) shows some improvements, especially in H$\alpha$ wings (to the sides of 656nm), H$\beta$ component positions (around 486nm; although, component depths are underestimated), and the flux curve below $\lambda 480$. Offset dipole configurations improve both fits. In the synthetic spectrum of \citet{kulebi2009} ($B_\text{d}=8.37$~MG, $i=6^\circ$, $a_z=0.31$), H$\alpha$ wings are corrected and the underestimation of the continuum below $\lambda 480$ nm is partially reduced. In our model ($B_\text{d}=12.28$~MG, $i=40^\circ$, $a_z=0.28$), the offset dipole improves H$\alpha$ wings, provides a better distribution of H$\beta$ components, and fits the bluer region of the spectrum slightly better. The magnetic-field parameters determined in this work are similar to those obtained by \citet{hardy2023} and those of \citet{amorim2023}, except for the displacement in the $z$ direction (see Table \ref{table1}).

%
\begin{figure}
\includegraphics[width=.49\textwidth]{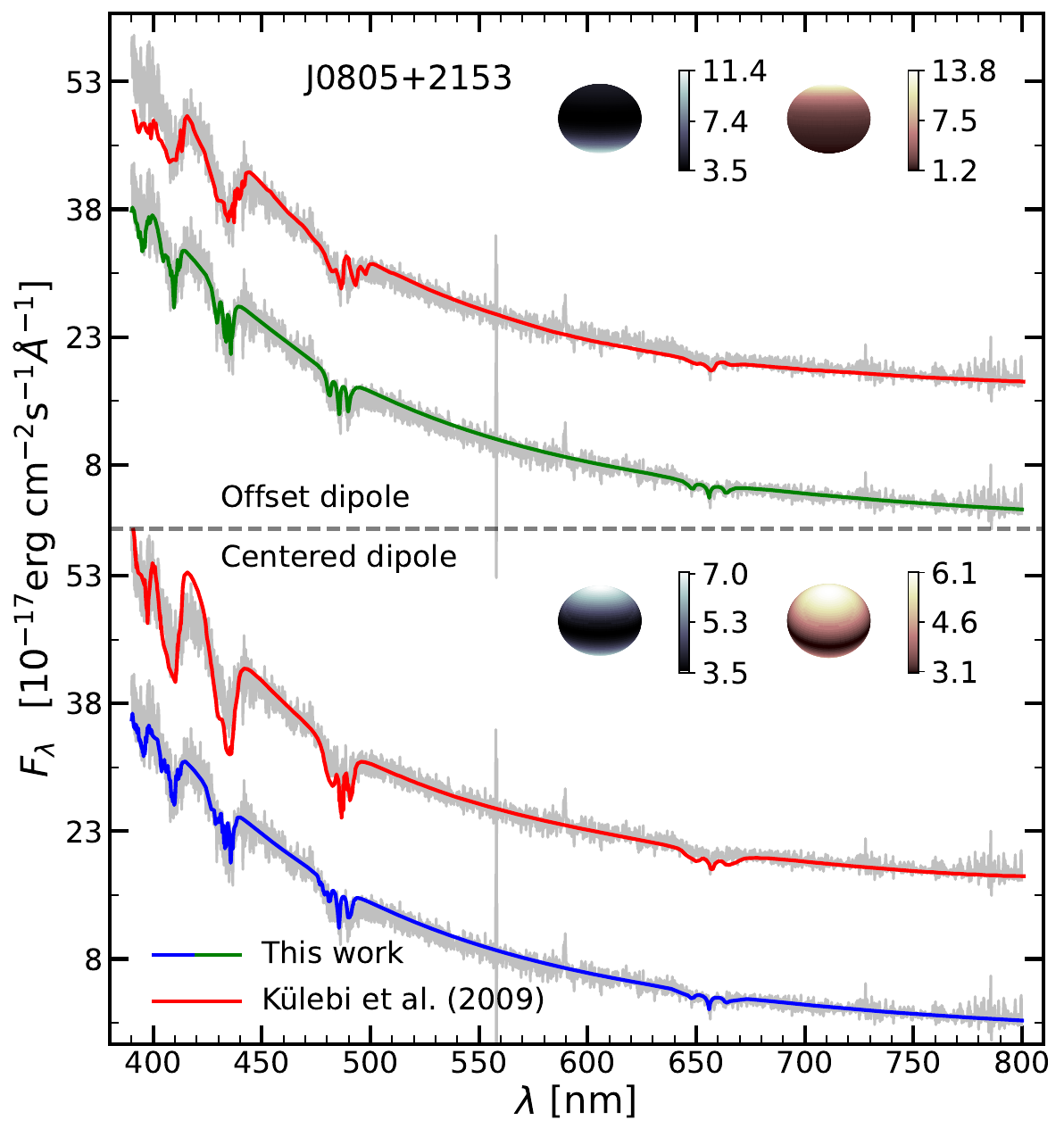}
\caption{Same as Fig. \ref{j1511}, but for J0805+2153. } 
\label{j0805}
\end{figure}   
%
\subsubsection{J0805+2153} 

This is the hottest object we studied. Its magnetic nature was first recognized by \citet{vanlandingham2005} from SLOAN DR3. They fit the SDSS spectrum with a centered magnetic dipole of $B_\text{d}=5$~MG, $i=60^\circ,$ and $T_\text{eff}=28000$~K. This object was also included in the surveys of \citet{eisenstein2006} and  \citet{kleinman2013}, with $T_{\text{eff}}=38211$~K and 37141~K, respectively. \citet{dufour2017} estimated its effective temperature to be relatively low (19713~K). Magnetic properties of this object were estimated in $3~$MG$~<B_\text{d}<7$~MG, $17^\circ~<i<87^\circ$~MG, and $-0.49~<a_z<0.3$ \citep{kulebi2009, kepler2013, amorim2023, hardy2023}.  Our optimal fit yields  $B_\text{d}=7$~MG, $i=85^\circ$, $a_z=-0.15$, and $T_\text{eff}=39000$~K. The variety of values obtained in different works for the inclination angle of the field and the shift of the dipole center (as in other objects analyzed here) suggests that these parameters (or the field geometry in general) are not well constrained.

Observed and calculated fluxes for J0805+2153 are displayed in Fig. \ref{j0805}. The centered dipole fit of \citet{kulebi2009} yields strong H$\beta$ components and features below $\lambda 450$, which are mostly corrected with a considerable shift of the dipole center ($a_z=0.39$) and a reduction of $B_\text{d}$ from 6.1~MG (centered) to 3.1~MG (offset). Our synthetic spectra with both centered and offset dipoles ($B_\text{d}\approx 7$~MG) reasonably reproduce most of the observed spectrum. 

\subsection{Additional spectral fits}
\label{s.additional}

\begin{figure}
\includegraphics[width=.50\textwidth]{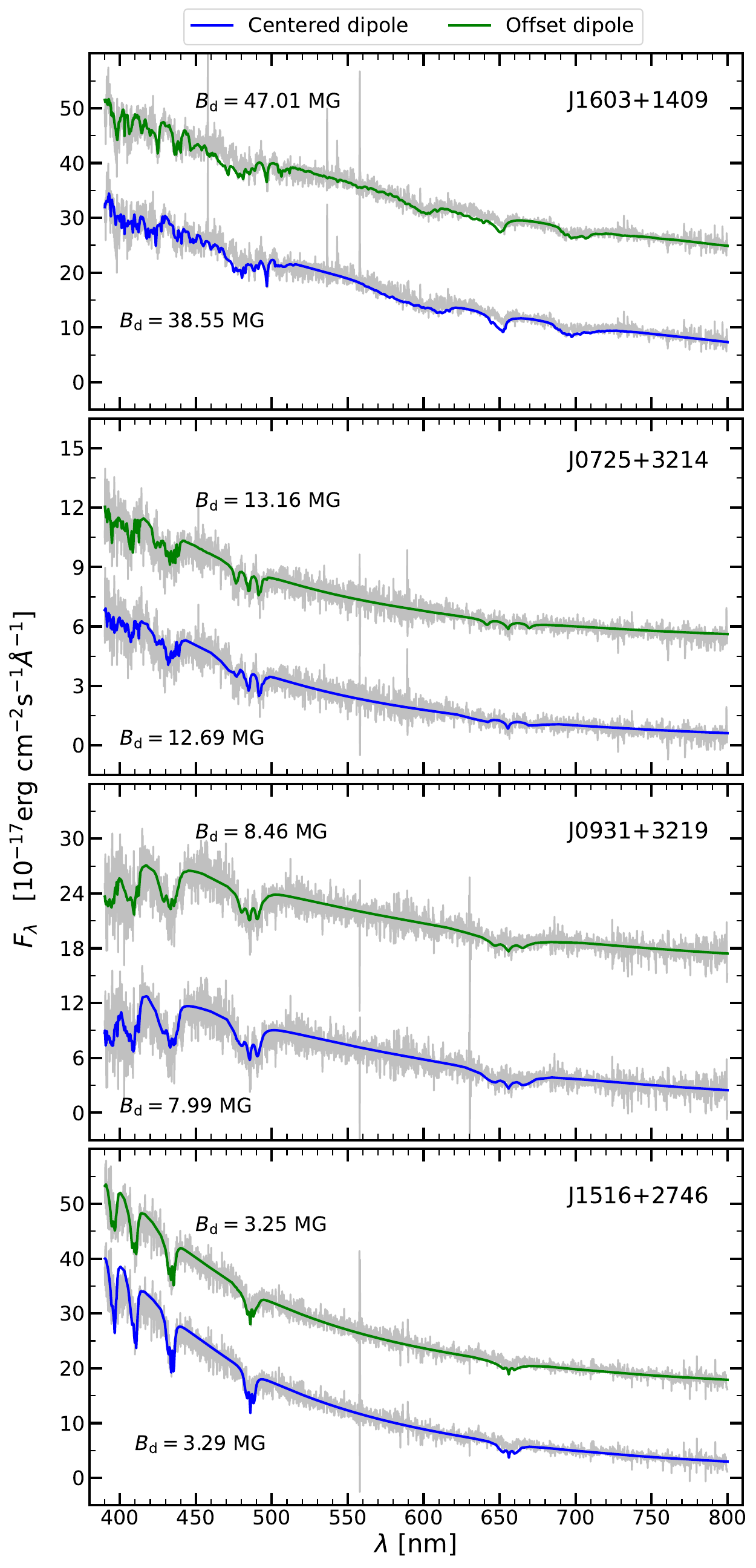}
\caption{Comparison between emerging flux of MWD stars from SLOAN Digital Sky Server (gray lines) and calculated by our code for a centered dipole model (blue lines) and offset dipole (green lines, displaced vertically for clarity). } 
\label{NKSpectra}
\end{figure}   
Figure \ref{NKSpectra} shows emerging fluxes (gray lines) of four MWD classified as white dwarfs by \citet{kleinman2013} from SDSS DR7 and identified as magnetic objects by \citet{kepler2013}. The magnetic nature of these MWDs was also analyzed by \citet{hardy2023} (except J1516+2746) and \citet{amorim2023}.  
Results in Fig. \ref{NKSpectra} are sorted upward by ascending magnetic dipole strength (comprising from $\sim 3 \, \text{MG}$ to $\sim 40 \, \text{MG}$), which shows the overall effect of field strength on white-dwarf spectra. In fact, as the field becomes stronger, hydrogen absorption lines are initially broadened (e.g., J1516+2746), split into different Zeeman components (J0725+3214), and finally mix together (especially members from high Balmer lines) when the field is strong enough (J1603+1409). Fig. \ref{NKSpectra} also displays our fits with centered (blue lines) and offset (green lines) dipoles compared with observed fluxes (no calculated emerging fluxes were found in the literature for these stars). We make a few comments for each object in the following sections.

\subsubsection{J1603+1409}

This object was reported as a short-period variable star by \citet{scholz2018} with an estimated variability period of $110\pm3$ min. Observed and calculated spectra are shown in the top of Fig. \ref{NKSpectra}. Magnetic strength in the stellar surface is high enough to cause line mixing below $520$~nm, whereas $\sigma_-$, $\pi$,  and $\sigma_+$  components of H$\alpha$ ($\approx\lambda 600$, $\lambda 650,$ and $\lambda 700$, respectively) become clearly separated. An overall agreement is obtained with a centered dipole of $B_\text{d}=38.55$~MG at a near pole-on view ($i=10^\circ$). A better fit results from a  strongly inclined ($i=70^\circ$) offset dipole ($a_z=0.15$) with a strength ($B_\text{d}=47.01$~MG) similar to those derived by \citet{amorim2023} and \citet{hardy2023} (Table \ref{table1}).

\subsubsection{J0725+3214}

This object represents an example of a warm MWD with moderate surface fields (Fig. \ref{NKSpectra}). Its SDSS spectrum is reasonably well represented with $B_\text{d}\approx 13$~MG and $T_\text{eff}=26000$~K. The low fitting error suggests that the field geometry is close to that of a dipole (Table \ref{table2}).  \citet{hardy2023} and \citet{amorim2023} found values slightly higher for $B_\text{d}$ and lower for $T_\text{eff}$ (Table \ref{table1}). 
On the other hand, the spectroscopic fitting of J0725+3214 with zero-field model atmospheres by \citet{kleinman2013} and \citet{dufour2017} gave appreciably higher effective temperatures, $T_\text{eff}=34711$~K and 28951~K, respectively (Fig. \ref{TBfits}).

\subsubsection{J0931+3219}

Strong absorption features are observed for this MWD, which can be reproduced with an effective temperature of 11000~K according to our synthetic spectrum (Fig. \ref{NKSpectra}). Values of $T_\text{eff}=11526$~K \citep{dufour2017} and $16248$~K \citep{kleinman2013} were derived from nonmagnetic models. Our flux fits revel moderate field strengths ($\ga 8$~MG), as that estimated through visual inspection of Zeeman splitting \citep{kepler2013}. A dipole shift along the $z$-axis reduces the fit error slightly. Somewhat higher $T_\text{eff}$ and $B_\text{d}$ values were found by \citet{hardy2023} and \citet{amorim2023}, as shown in Table \ref{table1} and Fig. \ref{TBfits}.

\subsubsection{J1516+2746} 

This object was studied as one of several hot white dwarfs from SDSS DR12 by \citet{bedard2020}, which inferred an effective temperature of $40581$~K by spectroscopic analysis. A higher value, $T_\text{eff}=67365$~K, was determined from color-color diagrams \citep{gentile2021}. Other estimates place $T_\text{eff}$ at lower values (Fig. \ref{TBfits}). A Zeeman pattern of H$\alpha$ and H$\beta$ suggests a mean field of $2.6<B<3.0$~MG \citep{kepler2013}. Using magnetic synthetic spectrum fitting, \citet{amorim2023} estimated $B_\text{d}=3.03$~MG and $T_\text{eff}=30000$~K. We obtained $B_\text{d}=3.25$~MG and $T_\text{eff}=33000$~K with an offset dipole configuration that reduces the fit error of our best centered dipole model (Fig. \ref{NKSpectra}).

%
\begin{table} 
\caption{Correlations on physical parameters derived from \citet{amorim2023} ($=A$), \citet{hardy2023} ($=H$) and this work ($=V$). Comparison between $H$ and $A$ corresponds to a common sample of 118 MWDs, most of them with moderate and low field strengths (115 stars with $B_\text{d}<50$~MG). 
}  \label{table3}
\begin{tabular}{ccccc}
\hline 
\hline
     &  $T_\text{eff}$ &  $B_\text{d}$  &  $i$  & $a_z$ \\
\hline
$A$--$V$  &   0.97  &  0.99  &   0.35   &   $-0.30$ \\     
$H$--$V$  &   0.99  &  0.98  &   0.04   &   $+0.41$ \\
$H$--$A$  &   0.69  &  0.96  &   0.15   &   $-0.23$ \\
\hline 
\end{tabular}
\end{table}

\subsection{Discussion}

Several results emerge from the spectral fits and comparisons previously analyzed. As stated in prior works, good fits to the observations are achieved for stars with low field strengths. Specifically, dipole field geometries are generally adequate to reproduce observed spectra in objects with mean field strengths lower than 50~MG, with offset dipoles usually providing the best fit. However, comparisons of physical parameters derived from different numerical codes reveal some heterogeneous results. In particular, reasonable agreement is found in the evaluation of the dipole strength, but significant differences appear in the derived values for the inclination of the dipole axis and its displacement from the stellar center.

Table \ref{table3} shows correlations between pairs of values derived from different studies for common groups of stars. Although the number of analyzed stars in the present work is small, some trends can be inferred by comparisons with results of \citet{amorim2023} and \citet{hardy2023}. Such trends are confirmed by comparisons between results from \citet{hardy2023} and \citet{amorim2023} for a sample of 118 stars studied in both works. Calculated correlation coefficients close to unity show a systematic agreement in determinations of $B_\text{d}$, indicating that the mean surface fields in spectrum fitting appear to be reasonably well determined.
On the contrary, derived values for $i$ and $a_z$ present very poor correlations.
This suggests that the dipole inclination and offset in numerical codes regulate the range of field strengths that effectively contribute to the spectral shape, rather than describing the exact geometric distribution of the field on the stellar surface.

On the other hand, spectrum fitting for magnetic white dwarfs with high field strengths becomes sensitive to magnetic-field effects on the ionization equilibrium in the atmosphere. A detailed search for the best spectrum fitting for the highly magnetic star SDSS J2247+1456, with and without field effects on the chemical equilibrium, showed a substantial change in the calculated $T_\text{eff}$, it being 1000~K higher in the former case (magnetic gas model). However, obtaining good spectrum fitting for highly magnetic white dwarfs is very difficult, likely because these stars have more complex field geometry than a dipole, or also because of the physical approximations used in the atmosphere models. Therefore, the search for the best fit usually yields several close solutions with $T_\text{eff}$, on the order of or greater than the change originating from the chemical model used. In any case, it is clear that the use of a field-free chemical model introduces a systematic error in spectral fitting, which mainly implies an underestimation of surface temperatures for highly magnetic objects.

\section{Conclusions}\label{s.conclusions}

We present a new code for synthetic spectrum calculation of pure-hydrogen MWD atmospheres. To our knowledge, our code is currently the only one capable of calculating a full solution to the radiative transfer equations, including the effects of the magnetic field on both opacities and abundances of atomic populations. We demonstrate that incorporating magnetic-field effects on the chemical equilibrium of the gas leads to a substantial increase in the temperature distribution in highly magnetized atmospheres. This impacts the determination of the effective temperature of highly magnetic white dwarfs and may consequently affect the characterization of other physical properties such as radius and mass. Although this effect on $T_\text{eff}$ is significant, its magnitude is on the same order as the current capability of magnetic synthetic spectrum models to reproduce observed spectra in these stars. Presently, effective temperature predictions for highly magnetic objects resulting from different atmosphere codes show discrepancies of a similar or even higher magnitude than changes arising from the use of a magnetic chemical model.

Comparison with results of other spectrum-fitting codes in a sample of MWDs using dipole geometry indicates reasonable agreement in the evaluation of the surface field strengths. However, there is notable disagreement in the identification of the dipole inclination relative to the observer and its offset from the stellar center, even for weak magnetic stars. On the other hand, differences found in the observed spectrum fits could indicate a magnetic field geometry that is more complex than a simple shifted dipole and uncertainties in the input physics of the atmosphere model. It remains to be seen whether more detailed agreement can be achieved when better continuum opacity data become available. Work in this regard is in progress.

\begin{acknowledgements}
This publication makes use of data from the Sloan Digital Sky Survey (SDSS) which is supported by the Alfred P. Sloan Foundation, the Heising-Simons Foundation, the National Science Foundation, and the Participating Institutions. 
\end{acknowledgements}

\bibliographystyle{aa}
\bibliography{MWDs}

\end{document}